\documentclass[a4paper,11pt]{article}
\pdfoutput=1
\usepackage{jheppub}
\usepackage[T1]{fontenc} 
\usepackage{slashed}
\usepackage[utf8x]{inputenc}
\usepackage{color}
\definecolor{red}{rgb}{1,0,0}
\usepackage[normalem]{ulem}
\usepackage{slashed}
\usepackage{bm}
\usepackage{bbold}
\definecolor{airforceblue}{rgb}{0.36, 0.54, 0.66}
\definecolor{steelblue}{rgb}{0.27, 0.51, 0.71}
\definecolor{amber}{rgb}{1.0, 0.49, 0.0}

\def\simg{{\ \lower-1.2pt\vbox{\hbox{\rlap{$>$}\lower6pt\vbox{\hbox{$\sim$}}}}\ }}
\def\siml{{\ \lower-1.2pt\vbox{\hbox{\rlap{$<$}\lower6pt\vbox{\hbox{$\sim$}}}}\ }}

\makeatletter \@addtoreset{equation}{section} \makeatother

\renewcommand{\[}{\begin{equation}}
\renewcommand{\]}{\end{equation}}
\def\beq{\begin{equation}}
\def\eeq{\end{equation}}
\newcommand{\be}{\begin{eqnarray}}
\newcommand{\ee}{\end{eqnarray}}

\def\({\left(}
\def\){\right)}

\newcommand{\LF}{\left(}
\newcommand{\RF}{\right)}
\newcommand{\LT}{\left[}
\newcommand{\RT}{\right]}

\title{Dark matter and Standard Model reheating from conformal GUT inflation}

\author[a]{S.~Biondini}
\author[a]{and K.~Sravan Kumar}

\affiliation[a]{Van Swinderen Institute, University of Groningen
 \\ 
Nijenborgh 4, NL-9747 AG Groningen, Netherlands}

\emailAdd{s.biondini@rug.nl}
\emailAdd{sravan.korumilli@rug.nl }

\abstract{Spontaneous breaking of conformal symmetry has been widely exploited in successful model building of both inflationary cosmology and particle physics phenomenology. Conformal Grand Unified Theory (CGUT) inflation provides the same scalar tilt and tensor-to-scalar ratio as of Starobinsky
and Higgs inflation. Moreover, it predicts a proton life time compatible with the current experimental bound.
 In this paper, we extend CGUT to account for the production of dark matter and the reheating of the Standard Model.  To this end, we introduce a hidden sector directly coupled to the inflaton, whereas the reheating of the visible sector is realized through a portal coupling between the dark particles and the Higgs boson. The masses and interactions of the dark particles and the Higgs boson are determined by the form of the conformal potential and the non-vanishing VEV of the inflaton. We provide benchmark points in the parameter space of the model that give the observed dark matter relic density and  reheating temperatures compatible with the Big Bang nucleosynthesis. }

\begin{document}
\maketitle
\flushbottom

\section{Introduction}

Inflationary cosmology stands today as an elegant and compelling explanation why our observable universe is so large, flat and homogeneous on large scales \cite{Starobinsky:1980te,Guth:1980zm,Linde:1981mu}. Quantum fluctuations during inflation provide the seeds for
structure formation, whereas the subsequent decay of the inflaton can trigger a hot thermal universe containing matter and radiation. According to the recent Planck data, single-field inflationary scenarios with a plateau-like (or Starobinsky-like) potentials are strongly favoured \cite{Ade:2015lrj,Akrami:2018odb}. Two models have gained significant attention: Starobinsky's $R+R^2$ inflation \cite{Starobinsky:1980te}, which is based on the modification of gravity induced by one-loop corrections from matter quantum fields \cite{Duff:1993wm}, and Higgs inflation \cite{Bezrukov:2008ej}, where the Standard Model (SM) Higgs boson is non-minimally coupled to gravity. These two inflationary models occupy a central spot in the $(n_s,r)$ plane with the predictions 
\begin{equation}
n_s = 1-\frac{2}{N},\qquad  r= \frac{12}{N^2}\,, 
\label{sweetspot}
\end{equation}
where $n_s$ is the scalar tilt, $r$ is the tensor-to-scalar ratio and $N$ is the number of e-foldings before the end of inflation. The observational bounds from the latest Planck data \cite{Akrami:2018odb} read $n_s=0.9649\pm 0.0042$ at $68\%$ CL and $r<0.064$ at $95\%$ CL.
The expressions (\ref{sweetspot}) are related to the shape of the potential during inflation which nearly coincides in both of these models  \cite{Kehagias:2013mya}.
The only way to distinguish between them is through the study of reheating \cite{Bezrukov:2011gp}.

In recent years, it has been realized that the Starobinsky-like inflation can also be induced through a spontaneous breaking of local scale invariance \cite{Kallosh:2013xya,Kallosh:2013hoa}.  This guiding principle is motivated by the approximately scale invariant power spectrum of the primordial cosmic fluctuations that suggests a role of scale symmetry in the inflationary model building. Throughout the paper, conformal symmetry is understood as a local scale symmetry, namely the action of the theory is invariant under space-time dependent transformations of the metric and the fields (cfr.~eq.~(\ref{ConformtrSU5})), at variance with a global scale symmetry or dilatation symmetry.\footnote{In the literature other expressions are used as well, for example local (global) scale invariance is dubbed local (global) Weyl symmetry, see e.g.~\cite{Ferreira:2016wem,Bars:2013yba}. Note that the conformal GUT model that we study here is different from the class of conformally invariant field theories associated with the conformal symmetry group (the extension of Poincar\'e group) at both classical and quantum level \cite{Armillis:2013wya,Low:2001bw}.} Global and local scale invariant models of inflation have been often adopted to explore a variety of different aspects: from inflationary perturbations, reheating, the SM electroweak symmetry breaking and the origin of the Higgs mass, as well as open issues in the SM like dark matter, baryogenesis and neutrino masses~\cite{GarciaBellido:2011de,Salvio:2017xul,Kannike:2015apa,Rinaldi:2014gha,Einhorn:2014gfa,Kannike:2014mia,Barrie:2016rnv,Tambalo:2016eqr,Farzinnia:2015fka,Ferreira:2016wem,Ferreira:2016vsc,Meissner:2006zh,Meissner:2008gj,Foot:2007iy,Chang:2007ki,Iso:2009ss,AlexanderNunneley:2010nw,Bars:2013yba,Carone:2013wla,Khoze:2013oga,Hambye:2013sna,Khoze:2014xha,Karam:2015jta,Khoze:2016zfi,Lewandowski:2017wov,Chankowski:2014fva,Davoudiasl:2014pya,Latosinski:2015pba,Demir:2013uja,Guo:2015lxa,Nayak:2017dwg,Croon:2019kpe,Foot:2007as,Foot:2007ay,Espinosa:2007qk,Englert:2013gz,Khoze:2013uia,Farzinnia:2013pga,Gabrielli:2013hma,Helmboldt:2016mpi,Oda:2017zul,Loebbert:2018xsd,Brdar:2019qut}.

Despite the spontaneous breaking of both global and confromal symmetry is related to a non-vanishing VEV of some of the scalar fields in the theory, there is a difference about the appearance of a dynamical Goldstone boson. In the case of globally scale invariant models, the symmetry breaking is accompanied with the corresponding massless Goldstone boson, which is a dynamical propagating field and has an impact in the early cosmology and phenomenology of a given model \cite{Wetterich:1987fm,GarciaBellido:2011de,Ferreira:2016wem,Ferreira:2016vsc}. In the case of conformally invariant models, the auxiliary scalar field, which is also called \textit{conformon} and comes with a wrong sign in the kinetic term, can be gauge fixed to an arbitrary value. The gauge fixing  can  be interpreted as a spontaneous breaking of conformal invariance due to existence of a classical field value for the conformon, however any dynamical property of the field is removed and no associated Goldstone boson emerges \cite{Kallosh:2013oma,Bars:2013yba,Bars:2012mt,SravanKumar:2018tgk}.  

 In this paper, we consider Conformal GUT (CGUT) inflation~\cite{SravanKumar:2018tgk}, that has been recently developed as a conformal extension of SU(5) GUT inflation~\cite{Starobinsky:1982ee,Shafi:1983bd}. Since the scale of inflation can be as high as $10^{14}$ GeV \cite{Martin:2015dha},
 it is quite appealing to consider the role of GUTs in the context of inflationary cosmology~\cite{Lyth:1998xn,Linde:2005ht,Mazumdar:2010sa,Hertzberg:2014sza,Martin:2015dha,Linde:2014nna,Elizalde:2014xva}. In CGUT inflation, conformal symmetry is introduced with two additional SU(5) singlet fields, one of them playing the role of the inflaton. 
 In this model, inflation occurs by spontaneous breaking of conformal and GUT symmetries. Conformal symmetry can be broken by gauge fixing the VEV of one of the singlet fields \cite{Kallosh:2013xya,Kallosh:2013hoa,Bars:2012mt,Bars:2013yba}, and it corresponds to a spontaneous breaking. In this work, we do not consider the possibility of an accompanying explicit breaking from an intrinsic mass scale, which can be generated through the renormalization procedure of the theory and the corresponding running couplings (scale anomaly) \cite{Wetterich:1987fm}.  We leave this interesting possibility for future investigations.
 Later on, a Coleman-Weinberg (CW) potential for the inflaton field is generated through the interactions with the GUT fields, where we take the couplings of the model as frozen and do not evolve them at different energy scales.
 One common aspect with GUT inflation is that the inflaton rolls down to a non-zero VEV \cite{Shafi:1983bd}, which offers a rich phenomenology and plays a crucial role in the dynamical generation of mass scales at the end of inflation~\cite{Lazarides:1991wu,Lazarides:1984pq}. However, there is an important difference with original GUTs models. In CGUT inflation the above VEV branch of the CW potential can be stretched to a Starobinsky-plateau leading to nearly the same predictions as (\ref{sweetspot}). 
 

The inflationary paradigm  encompasses two aspects through the inflaton field. First, the inflaton perturbations explains the observed nearly scale invariant density fluctuations in the CMB. Second, the inflaton may be responsible for particle production through its decays during the reheating stage at the end of inflation. This way the present matter and radiation content in the universe is originated from a single field. Since dark matter appears to be the dominant matter component already at the time of CMB,
it is natural to think of a strong connection between the inflaton field and a dark sector. As noted, the great desert that lies  between the scale of inflation ($\sim 10^{14}$ GeV) and the electroweak scale can be an artefact of some hidden sector we have not yet observed.

Motivated by a win-win interplay between particle physics and cosmology, we further develop CGUT inflation to account for a viable dark matter particle and the electroweak symmetry breaking. The first aspect is realized by letting the inflaton decay into a dark sector, made of fermions and scalar particles. We assume the dark fermion to be an inert and stable particle that accounts for the present-day relic density. The masses of the particles in the hidden sector are induced by the spontaneous breaking of conformal and GUT symmetry, which are at the origin of the inflaton VEV as well. 
The second aspect is addressed through a portal coupling between the Higgs boson and the dark scalar. When the latter reaches its non-vanishing VEV, it generates a mass scale for the Higgs boson that triggers the electroweak symmetry breaking. The dark scalar can decay into Higgs bosons pairs and induce the reheating of the SM sector. This latter aspect of our framework is similar to other recent implementations for a SM reheating from hidden sectors \cite{Tenkanen:2016jic,Berlin:2016vnh,Paul:2018njm}. In our construction,  conformal symmetry breaking is responsible for the generation of the relevant scales, from inflation down to the electroweak scale. We note by passing that a cosmological constant can be generated by conformal symmetry breaking as considered in refs.~\cite{Sadeghi:2015bxy,Oda:2018zth,Bloch:2019bvc}. 

The paper is organized as follows. In section~\ref{sec:fund_symm} we introduce the different scalar potentials appearing in cosmology and particle physics and relevant to our framework. In section~\ref{sec:conf_infl}, the CGUT framework is briefly reviewed. We introduce the masses and couplings generation in the dark sector in section~\ref{sec:Infl_matter}, together with a discussion about the constraints on the model parameters.  The non-thermal production of dark matter and the reheating of the SM from the hidden sector is discussed in section~\ref{sec:numerical_final}, whereas conclusions are offered in section \ref{sec:conclusion}.

\section{Fundamental potentials in cosmology and particle physics}
\label{sec:fund_symm}

In this section we discuss how the scalar field potentials, that frequently occur in our current understanding of inflationary cosmology and particle physics, can be related to conformal symmetry. 
Let us start with particle physics. The Higgs boson is the only fundamental scalar of the SM
and it is responsible for the generation of the fermions and gauge bosons masses via spontaneous breaking of the gauge symmetry. This is implemented with the well-known  Brout–Englert–Higgs mechanism \cite{Higgs:1964pj,Higgs:1964ia,Englert:1964et}, which assumes a negative mass term compatible with the gauge symmetry and that induces a non-trivial minimum of the potential. The Higgs potential reads
\begin{equation}
V_H=-\mu_H^2 H^\dagger H + \lambda_H (H^\dagger H)^2 \, ,     
\label{higgs_0_pot}
\end{equation}
where $H$ is the Higgs doublet, $\mu_H^2>0$ is the Higgs mass term and $\lambda_H$ is the Higgs self-coupling. The condition for the minimum brings to a VEV $v^2_H=\mu^2_H/\lambda_H$ and the physical Higgs mass is $m^2_H =2 \lambda_H v_H^2=2 \mu_H^2$. The negative quadratic term in (\ref{higgs_0_pot}) destabilizes the potential at the origin. Such a solution for the electroweak symmetry breaking predicts that the Higgs quartic coupling, as inferred from the measurement of the Higgs boson mass measurement, also determines the strength of self-interactions of the Higgs boson. Checking this prediction is very important to confirm that the electroweak symmetry breaking is induced by the potential in (\ref{higgs_0_pot}). However, current measurements at the LHC  leave still room for deviations induced by new physics and the details of the symmetry breaking can be perhaps probed at the LHC upgrades, muon and linear  collider  facilities \cite{Gupta:2013zza,Chiesa:2020awd,Baer:2013cma,Fuks:2017zkg,Cepeda:2019klc}.

Alternatively, we may start with a scale invariant theory and set $\mu_H=0$ in (\ref{higgs_0_pot}).
The idea of an underlying conformal symmetry is attractive since it allows to abandon the
arbitrary negative mass parameter for the Higgs field.
However, we can still achieve spontaneous breaking of symmetries through radiative corrections and let the scalar field acquire a VEV. This is the well-known Coleman-Weinberg (CW) mechanism~\cite{Coleman:1973jx}, where radiative
corrections to the Higgs self-coupling destabilize the Higgs potential at the origin. Despite the very minimal conformal SM 
is not capable of explaining the observed particles masses~\cite{Coleman:1973jx}\footnote{In ref.~\cite{Coleman:1973jx} the top quark was not taken into account because it had not yet been discovered at that time,
which led the authors to the conclusion that a stable radiatively generated minimum can be attained. However, the predicted mass of the Higgs boson was too low. After including the top quark, it was shown that there is no stable minimum due to the large contribution of top-quark Yukawa to the Higgs self-coupling running.}, this framework can be successful when some
field content that goes beyond the SM is included. The main point is that the Higgs mass term $\mu_H$ is reinterpreted in terms of a vacuum expectation value of a new scalar, coupled to the SM via a Higgs portal interaction.  In this respect, several extensions of SM were considered, where  $\mu_H$ is generated through a portal coupling with particles of a hidden sector \cite{Meissner:2006zh,Foot:2007as,Espinosa:2007qk,Foot:2007iy,AlexanderNunneley:2010nw,Englert:2013gz,Khoze:2013uia,Bars:2013yba,Farzinnia:2013pga,Gabrielli:2013hma,Lewandowski:2017wov,Chankowski:2014fva}.  Since the CW potential will appear in our framework as well, we give its parametric form for a generic scalar field $\phi$ and the radiatively generated VEV $v_{\hbox{\tiny CW}}$ as follows
\begin{equation}
    V_{\hbox{\tiny CW}}   \simeq  \, A_{\hbox{\tiny CW}} \phi^4 \LT \ln\LF \frac{\phi}{v_{\hbox{\tiny CW}}} \RF -\frac{1}{4} \RT  + A_{\hbox{\tiny CW}} \frac{ v_{\hbox{\tiny CW}}^4}{4} \, ,
    \label{CW_0_pot}
\end{equation}
where $ A_{\hbox{\tiny CW}}$ is a dimensionless quantity. 

On the early universe cosmology side, single-field inflationary scenarios with plateau-like (Starobinsky-like) potentials 
\begin{equation}
V_{\hbox{\tiny S}}  = A_{\hbox{\tiny S}}  \LF 1-e^{-\sqrt{\frac{2}{3}}\frac{\phi}{M_{\hbox{\tiny P}}}} \RF^2\,, 
\label{staropot}
\end{equation}
 are highly successful with respect to cosmological data \cite{Kehagias:2013mya,Akrami:2018odb}.
In (\ref{staropot}) $\phi$ is the inflaton field, $A_{\hbox{\tiny S}}$ is a dimensionful quantity and $M_{\hbox{\tiny P}}$ is the reduced Planck mass. The above potential has been realized in various frameworks. The Starobinsky model based on $R+R^2$ modification of gravity in the Einstein frame has exactly the same form as the potential (\ref{staropot}) \cite{Starobinsky:1980te,Kehagias:2013mya}. The success of the Starobinsky model with respect to the CMB data can be heuristically understood as the $R^2$ term is scale invariant. 
Interestingly, the potential (\ref{staropot}) can be also obtained with a two-field model with conformal symmetry, which can be spontaneously broken by gauge fixing one of the field to its VEV \cite{Kallosh:2013xya,Kallosh:2013hoa}. Also Higgs inflation leads to the same shape of the potential (\ref{staropot}) in the scale invariant regime, namely when the Higgs field is far away from electroweak vacuum so that the mass term can be neglected. However, after inflation the Higgs field acquires a non-zero VEV leading to mass scales of SM degrees of freedom \cite{Bezrukov:2007ep}. 

In this work we develop both successful inflation and post-inflationary particle physics in one single framework by invoking conformal symmetry as our guiding principle. We start with CGUT inflation as implemented in ref.~\cite{SravanKumar:2018tgk} and we extend it in order to discuss post-inflationary physics that involves both a dark sector and the visible sector. In our framework all three potential forms in (\ref{higgs_0_pot}), (\ref{CW_0_pot}) and (\ref{staropot}) are realized as originated by symmetry breaking patterns.

\section{Conformal GUT inflation}
\label{sec:conf_infl}
Inflationary cosmology is an effective field theory (EFT) for energy scales much below the Planck scale or the so-called regime of quantum gravity. Here, we consider an EFT dubbed as CGUT which  describes the physics from GUT energy scales $\LF \sim 10^{16}\textrm{GeV} \RF$ down to smaller scales.
The framework of CGUT inflation in SU(5) can be described by the following action \cite{SravanKumar:2018tgk} 
\begin{equation}
\begin{split}S_{\hbox{\tiny CGUT}}= & \int d^{4}x\,\sqrt{-g}\Bigg[\left(\chi^{2}-\phi^{2}-\textrm{Tr}\Sigma^{2}\right)\frac{R}{12}-\frac{1}{2}\left(\partial\phi\right)\left(\partial\phi\right)+\frac{1}{2}\left(\partial\chi\right)\left(\partial\chi\right)\\
 & -\frac{1}{2}\text{Tr}\left[\left(D^{\mu}\Sigma\right)^{\dagger}\left(D_{\mu}\Sigma\right)\right]-\frac{1}{4}\text{Tr}\left(F_{\mu\nu}F^{\mu\nu}\right)-V \Bigg]\,,
\end{split}
\label{CFTSU(5)}
\end{equation}
where $\LF \phi,\,\chi \RF$ are real singlets of SU(5) conformally coupled to GUT Higgs field $\Sigma$, which belongs to the adjoint representation of SU(5), and $R$ is the Ricci scalar. 
The covariant derivative is defined by $D_{\mu}\Sigma=\partial_{\mu}\Sigma-ig\left[A_\mu,\,\Sigma\right]$,
$A_{\mu}$ are the 24 massless Yang-Mills fields with field strength given by $F_{\mu\nu}\equiv\nabla_{[\mu}A_{\nu]}-ig\left[A_{\mu},\,A_{\nu}\right]$. Here, the 
field $\phi$ is responsible for inflation when the conformal symmetry is broken by the field $\chi$ as we will explain shortly.

The tree level potential in (\ref{CFTSU(5)}), that accounts for the interactions between the inflaton and the other fields, can be split into two contributions as follows
\begin{eqnarray}
V= V\left(\phi,\,\chi,\,\Sigma\right) + V(\phi,\hbox{\scriptsize SM},\hbox{\scriptsize DS}) \, ,
\label{pot_infl_reheat}
\end{eqnarray}
where $V(\phi,\hbox{\scriptsize SM},\hbox{\scriptsize DS})$ includes interactions of the inflaton field with the dark sector (DS), which are negligible during inflation (see discussion in section~\ref{sec:Infl_matter}). 
Its expression is given later in section~\ref{sec:Infl_matter} (cfr.~eq.~(\ref{bosf})), and it is (i) conformal invariant and (ii) responsible for the reheating process that will be discussed in section~\ref{sec:numerical_final}. 
The first term in eq.~(\ref{pot_infl_reheat}), which is instead responsible for the inflationary dynamics, reads 
\begin{equation}
V\left(\phi,\,\chi,\,\Sigma\right)=\frac{a}{4}\left(\textrm{Tr}\Sigma^{2}\right)^{2}+\frac{b}{2}\textrm{Tr}\Sigma^{4}-\frac{\lambda_{2}}{2}\phi^{2}\textrm{Tr}\Sigma^{2}f\left(\frac{\phi}{\chi}\right)+\frac{\lambda_{1}}{4}\phi^{4}f^{2}\left(\frac{\phi}{\chi}\right) \, ,
\label{potCFTSU5}
\end{equation}
where $a \approx b \approx g^{2}$, $g$ is the gauge coupling of the GUT group with  $g^{2}\simeq 0.3$ as obtained from the fine structure constant $\alpha_G = g^2/(4\pi) \simeq 1/40$ \cite{Rehman:2008qs}. The SU(5) group contains another scalar which is the fundamental Higgs field $H_{5}$, which comprises the colour-triplet Higgs and the SM Higgs doublet. In eq.~(\ref{potCFTSU5}) we assume the coupling between the field $\phi$ and $H_5$ to be negligible in comparison with its coupling to the adjoint field $\Sigma$~\cite{Esposito:1992xf,SravanKumar:2018tgk}. We further assume the couplings between $\Sigma$ and $H_5$ to be very small and do not play a role in the inflationary dynamics \cite{Rehman:2008qs} (we elaborate more on this point in section~\ref{sec:proton_decay}).

As consistent with conformal symmetry, there is no mass scale in eq.~(\ref{potCFTSU5}). 
The main feature of the potential is the appearance of field-dependent couplings that depend on the ratio of the fields $(\phi,\chi)$ through $f\LF \phi/\chi \RF$~\cite{Kallosh:2013xya,Kallosh:2013hoa,Bars:2013yba}.
Given the tree-level potential in (\ref{potCFTSU5}), the action (\ref{CFTSU(5)})
is then conformally invariant under the following transformations
\begin{equation}
g_{\mu\nu}\to\Omega^{2}\left(x\right)g_{\mu\nu}\quad,\quad\chi\to\Omega^{-1}(x)\chi\quad,\quad\phi\to\Omega^{-1}\left(x\right)\phi\quad,\quad\Sigma\to\Omega^{-1}\left(x\right)\Sigma\,.\label{ConformtrSU5}
\end{equation}
In principle any generic function $f\LF\phi/ \chi \RF$ is allowed with respect to conformal invariance. In the context of a successful inflation, the following choice  was found to be useful~\cite{Kallosh:2013hoa,SravanKumar:2018tgk} 
\begin{equation}
f\left(\frac{\phi}{\chi}\right)=\left(1-\frac{\phi^{2}}{\chi^{2}}\right)\,.\label{fPchi}
\end{equation}
In CGUT the first symmetry breaking pattern is performed by letting the field $\chi$ acquire a constant value $\chi = \sqrt{6}M$, where $M$ is the mass scale associated with the conformal symmetry breaking. As discussed and implemented in refs.~\cite{Kallosh:2013xya,Kallosh:2013hoa,Bars:2013yba,SravanKumar:2018tgk}, the conformon field $\chi$ is gauged fixed to a constant value, also called as \textit{c-gauge} in the context of SUGRA frameworks \cite{Bars:2013yba}, and there is no associated dynamical massless degree of freedom.\footnote{It was explicitly shown in ref.~\cite{Jackiw:2014koa} that the conformal symmetry we discuss here has no associated conserved current and thus has no dynamical role once we gauge fix the field.} The situation is different from the case of a global scale symmetry breaking, where the associated Goldston boson, also called dilaton, has an impact on the early cosmology \cite{GarciaBellido:2011de,Ferreira:2016vsc,Ferreira:2016wem}.

Next, we consider the GUT symmetry breaking $\textrm{SU}(5)\to\textrm{SU}(3)_{c}\times\textrm{SU}(2)_{L}\times\textrm{U}(1)_{Y}$
by\footnote{In refs.~\cite{Shafi:1983bd,Rehman:2008qs} the notation $\langle \Sigma \rangle$ is instead used in eq.~(\ref{GUTfieldVEV-1}) and it stands for a particular direction in the vector space of SU(5).}
\begin{equation}
\Sigma=\sqrt{\frac{1}{15}}\sigma \,  \textrm{diag}\left(1,\,1,\,1,-\frac{3}{2},\,-\frac{3}{2}\right)\,.\label{GUTfieldVEV-1}
\end{equation}
Assuming $\lambda_{1}\ll\lambda_{2}\ll a,\,b$ and due to the coupling
$-\frac{\lambda_{2}}{2}\phi^{2}\textrm{Tr}\Sigma^{2}f\left(\frac{\phi}{\sqrt{6}M}\right)$,
the GUT field $\sigma$ reaches its local field dependent minimum
that reads 

\begin{equation}
\sigma^{2}=\frac{2}{\lambda_{c}}\lambda_{2}\phi^{2}f\left(\frac{\phi}{\sqrt{6}M}\right)\, ,
\label{sigma2field}
\end{equation}
where $\lambda_c= a+7b/15$~\cite{Rehman:2008qs}. 
As we can see from eq.~(\ref{sigma2field}), the field $\sigma$ continues to follow the behaviour of the
field $\phi$. The tree level potential for the $\left(\phi,\,\sigma\right)$
sector is given by

\begin{equation}
V\left(\phi, \sigma \right) =\left[\frac{\lambda_{c}}{16}\sigma^{4}-\frac{\lambda_{2}}{4}\sigma^{2}\phi^{2}f\left(\frac{\phi}{\sqrt{6}M}\right)+\frac{\lambda_{1}}{4}\phi^{4}f^{2}\left(\frac{\phi}{\sqrt{6}M}\right)\right]\,,\label{treelevel3fieldpot}
\end{equation}
and the effective potential for the
inflaton field $\phi$ due to the radiative corrections becomes \cite{SravanKumar:2018tgk}
\begin{equation}
V_{\hbox{\scriptsize eff}}\left(\phi\right)= A\frac{M^4}{m_P^4}\phi^{4}f^{2}\left(\frac{\phi}{\sqrt{6}M}\right)\left(\ln\left(\frac{\phi M \sqrt{f\left(\frac{\phi}{\sqrt{6}M}\right)}}{v_\phi M_{\rm P}}\right)-\frac{1}{4}\right)+\frac{Av_\phi^{4}}{4}\,,
\label{vef-1}
\end{equation}
where $A \simeq \lambda_{2}^{2}/(16\pi^{2})$, $\langle \phi \rangle \equiv v_\phi$ is the VEV of the inflation, the counterterm has been fixed to $\delta V=\frac{\delta{\lambda}_{2}}{4}\sigma^{2}\phi^{2}f^{2}\left(\frac{\phi}{\sqrt{6}M}\right)$,
the normalization constant is such that $V_{\hbox{\scriptsize eff}}\left(\phi=v_\phi\right)=0$
and the corresponding vacuum energy density is $V_{0} \equiv V_{\hbox{\scriptsize eff}}\left(\phi=0\right)  =A v_\phi^{4}/4$.
In CGUT we can generate the Planck mass dynamically by fixing
\begin{equation}
v_\phi=\sqrt{6}\,  M_{\rm P} \left( \frac{1}{\gamma^2} - 1 \right)^{\frac{1}{2}} \,.\label{phiVev}
\end{equation}
where $M_{\text{P}}=2.43 \times 10^{18}$ GeV is the reduced Planck mass and $\gamma \equiv M_{\rm P}/M$, which is smaller than unity by definition in our model. We notice that the inflaton effective potential in eq.~(\ref{vef-1}) has the form of a CW potential in eq.~(\ref{CW_0_pot}). To summarize, inserting (\ref{sigma2field}) and (\ref{treelevel3fieldpot}) in the original action (\ref{CFTSU(5)}), we obtain a quantum effective action for the inflaton field $\phi$ that reads
\begin{equation}
    S_{\phi}=\int d^4x\sqrt{-g}\Bigg[\LF 6M^2-\phi^2 \RF \frac{R}{12}-\frac{1}{2}(\partial\phi)(\partial\phi) -V_{\hbox{\scriptsize eff}}(\phi)\Bigg]\,.
    \label{effphi}
\end{equation}
\begin{figure}[t!]
    \centering
    \includegraphics[scale=0.93]{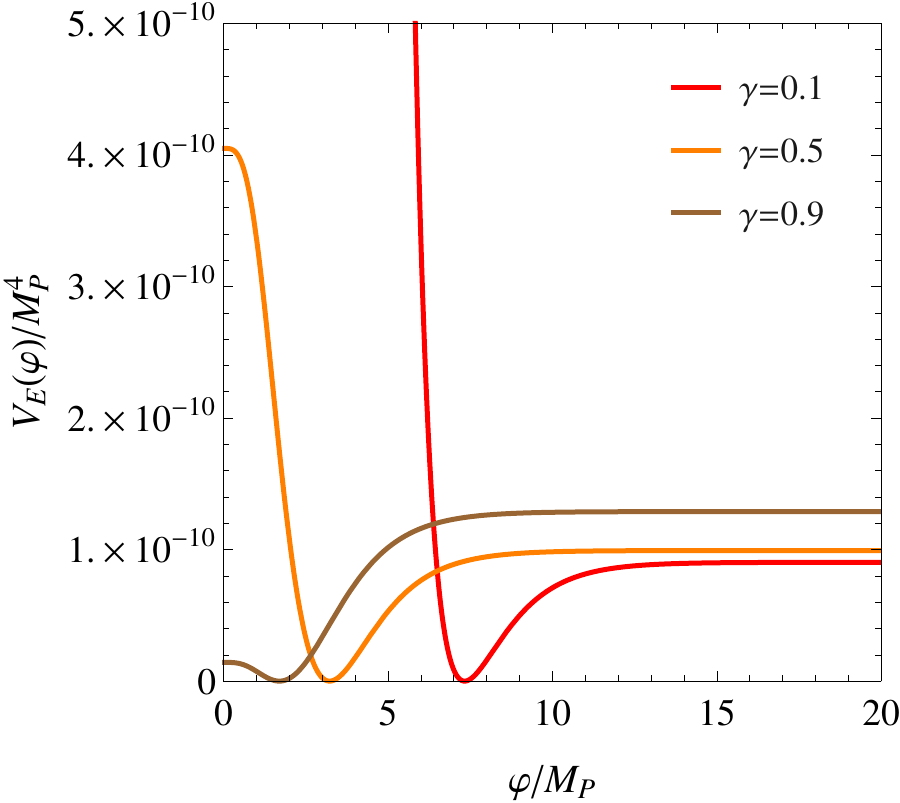}
    \caption{We plot the potential $V_E(\varphi)$ for $A= 5\times 10^{-12}$ from Table 1 of \cite{SravanKumar:2018tgk} for different values of $\gamma$. We can notice that the potential shape reaches a Starobinsky Plateau in the limit when $\phi\to \sqrt{6}M$ or $\varphi\gg \sqrt{6}M$.}
    \label{fig:potential_platau}
\end{figure}

In order to clearly see  the Starobinsky-like  inflation in this theory, we perform the conformal transformation of the action (\ref{effphi}) and canonically normalize the scalar field as $\phi=\sqrt{6}M\tanh\left(\frac{\varphi}{\sqrt{6}M_{\textrm P}}\right)$. Thus, we obtain a minimally coupled scalar field $\varphi$ in the Einstein frame  with the following potential\footnote{Note that when we go to the Einstein frame we must do a rescaling of mass scales with the conformal factor as $v_\phi^2 \to \frac{6v_\phi^2 M_{\textrm{P}}^2}{\LF 6M^2-\phi^2 \RF}$. Since at the end of inflation  $6M^2-v_\phi^2=6M_{\rm P}^2$, we can clearly see the VEV of the inflaton field remains the same in Jordan and Einstein frames.} 

\begin{equation}
V_{E}\left(\varphi\right)=36AM^4\tanh^{4}\left(\frac{\varphi}{\sqrt{6}M_{\rm P}}\right)\left(\ln\left(\frac{\sqrt{6}M\tanh\left(\frac{\varphi}{\sqrt{6}M_{\textrm P}}\right)}{v_\phi}\right)-\frac{1}{4}\right)+\frac{Av_\phi^{4}}{4}\,,\label{varphipot}
\end{equation}
where
the corresponding VEV of the canonically normalized scalar field is $\langle \varphi \rangle \equiv v_\varphi=\sqrt{6}\arctan\left(\frac{v_\phi}{\sqrt{6}M}\right)$. 
We can now easily notice that for $\varphi \gg \sqrt{6}M_{\textrm P}$ (i.e., $\phi\to \sqrt{6}M$), the potential (\ref{varphipot}) reaches a plateau.
During the inflationary regime, the shape of the potential is nearly the same as the Starobinsky and Higgs inflation as shown in Fig.~\ref{fig:potential_platau}.

The key inflationary predictions ($n_s,\,r$) in this model were computed up to leading order in the slow-roll approximation~\cite{SravanKumar:2018tgk} and turn out to be as in (\ref{sweetspot}), namely 
they are the same as those of Starobinsky inflation. However, there is a crucial difference since the inflaton reaches a non-zero VEV at the end of inflation, which can be seen in Fig.~\ref{fig:potential_platau}. 
We can fix the coupling $\lambda_2$, that enters the amplitude of the potential, by using the CMB constraint  on the scalar power spectrum  \cite{SravanKumar:2018tgk,Akrami:2018odb}  
\begin{equation}
    \mathcal{P}_\mathcal{R} = \frac{H_\ast^2}{6\pi^2M_{\textrm{P}}^2}N^2\approx 2.2\times 10^{-9} \,,
    \label{prl}
\end{equation}
where the value of Hubble parameter during inflation can be read from Fig.~\ref{fig:potential_platau} as $H_\ast \approx \sqrt{\frac{1}{3M_{\textrm P}^2}V_E^*}\approx 1.5\times 10^{13}\,\textrm{GeV} $. Numerical estimates of $\lambda_2$ using eq.~(\ref{prl}) for $\gamma < 0.9$ are found to be\footnote{We read the average value of $\lambda_2$ taking $A\sim 5\times 10^{-12}$ from Table I of \cite{SravanKumar:2018tgk} since there is very mild dependence on $\gamma$.} \cite{SravanKumar:2018tgk} 
\begin{equation}
    \lambda_2^2 \approx 8\times 10^{-10}\,. 
    \label{l2}
\end{equation}
We notice that the value of $\lambda_2$ is nearly the same for any VEV of the inflaton field as studied in~\cite{SravanKumar:2018tgk}. One may understand from Fig.~\ref{fig:potential_platau} that the shape of the potential during inflation remains nearly the same for any value of the inflaton VEV. This implies we can compute the inflaton mass at the end of inflation in terms of $\gamma$ as follows
\begin{equation}
M_{\Phi}= \left. \sqrt{V^E_{\varphi,\varphi}} \right|_{\varphi=\langle \varphi \rangle} \simeq 2\times 10^{-6} v_\phi \simeq 5 \times 10^{-6}  \frac{M_{\textrm P}}{\gamma} \left( 1-\gamma^2\right)^{1/2} \, .
\end{equation}
From the last expression we see that the mass of the inflaton increases for smaller values of $\gamma$. We avoid $\gamma \ll 1$ in order not to have an inflaton mass too close to the Planck mass according to the point of view of an effective theory for inflation \cite{Cheung:2007st}. 
In this paper we take  $\gamma=0.1$ as the lowest value,
for which we get $M_\Phi \simeq 1.2 \times 10^{14}$ \text{GeV}.
\begin{figure}[t!]
    \centering
    \includegraphics[scale=0.65]{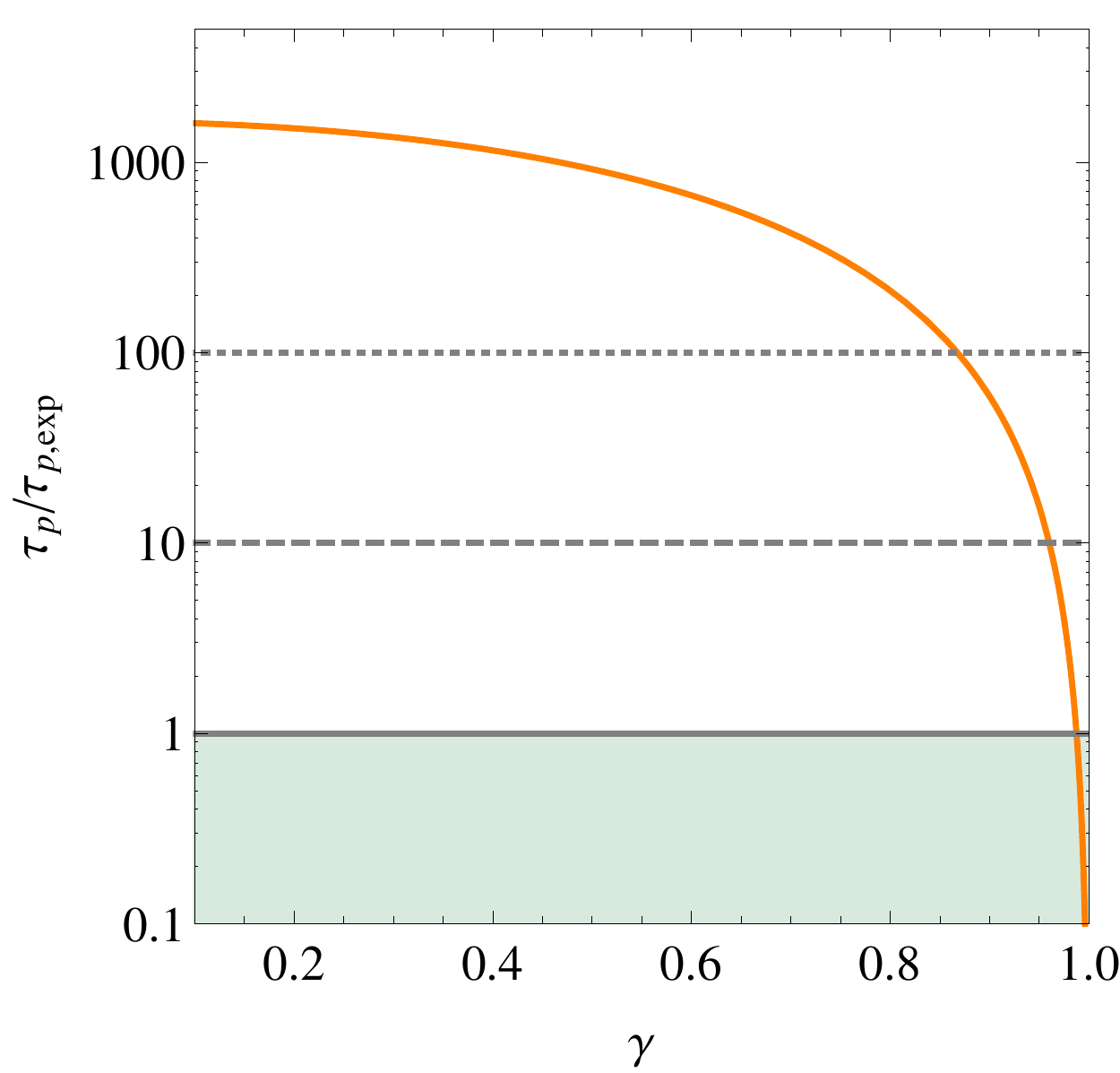}
    \caption{The ratio between the proton lifetime  predicted in the CGUT model and the observed lower bound is shown with a solid orange line. The forbidden region is indicated with the gray shaded area and gives $\gamma \leq 0.97$. The gray dashed and dotted lines indicate 10 and 100 times the experimental bound lower bound.}
    \label{fig:protondecay}
\end{figure}
\subsection{Proton decay}
\label{sec:proton_decay}
As one can read off eq.~(\ref{sigma2field}), the GUT field reaches a VEV through the inflaton field. Accordingly the GUT gauge bosons acquire a mass and they can mediate the proton decay\footnote{Since we assume the couplings between $\Sigma$ and $H_5$ to be negligible, we do not consider the color-triplet Higgs and its effect in the proton decay process. We take the proton decay as mediated by $X, Y$ gauge bosons in this minimal framework.} (they are often labelled with $X$ and $Y$). As a general feature of GUT models, the larger the gauge boson masses the longer the proton life time. In this particular model, the VEV of the inflaton field $\phi$, the GUT gauge boson masses, and consequently the proton life time, depend on one free parameter $M$ that we trade for $\gamma$. The mass of the gauge bosons depends on the VEV of the GUT field as follows~\cite{Rehman:2008qs}
\begin{equation}
    M_X = \sqrt{\frac{5}{3}} \frac{g v_\sigma}{2} \simeq\sqrt{{5} \lambda_2\LF 1-\gamma^2 \RF} M_{\textrm P}\,.
    \label{mx}
\end{equation}
where the second equality applies to our model, $\langle \sigma \rangle \equiv v_\sigma$ and we take $\lambda_c \simeq g^2$. 
The proton life time in this model can be computed as\footnote{In \cite{SravanKumar:2018tgk} the proton life time was computed by approximating the gauge boson mass as $M_X\sim 2 V_0^{1/4}$. We have found that this approximation is incorrect in CGUT inflation as it is only valid in the case of standard GUT inflation \cite{Rehman:2008qs}. We have provided the correct expression in eq.~(\ref{mx}) by including accordingly the dependence of $X$ boson mass on the parameter $\gamma$. This alters the estimates of proton life time as compared to \cite{SravanKumar:2018tgk}, however the model still gives prediction much above the present lower bound as shown in Fig.~\ref{fig:protondecay}.  }  \cite{Nath:2006ut,SravanKumar:2018tgk}  
\begin{equation}
\tau_{p}\approx \frac{M_X^4}{\alpha_G^2m_{p}^5} \approx 3.2 \times 10^{-5}\frac{M_{\rm P}^4}{m_{p}^{5}}  \LF 1-\gamma^2 \RF^2 \approx 2.6 \times 10^{37}  (1-\gamma^2)^2 \,  \hbox{yrs} \,,
\label{prlifetime}
\end{equation}
where $m_{p}$ is the proton mass and we used the estimate of $\lambda_2$ from eq.~(\ref{l2}).  The current lower bound on proton life time is given
by $\tau_{p,\hbox{\scriptsize exp}}>1.6\times10^{34}$ years \cite{Nishino:2009aa,Miura:2016krn} and it allows for $\gamma \lesssim 0.97 $
 as one may see in Fig.~\ref{fig:protondecay}. However, in order not to make the Planck scale completely approaching the conformal  scale, we consider $\gamma < 0.9$. In the remaining of this work, we take $0.1 \leq \gamma \leq 0.9$ for which $\tau_p/\tau_{p,\hbox{\scriptsize exp}} \simeq 1600$ and  $\tau_p/\tau_{p,\hbox{\scriptsize exp}} \simeq 60$ respectively.

In standard GUTs, $\Sigma$-$H_5$ interactions are needed in order to give the Higgs doublet a mass term to trigger the electroweak symmetry breaking. At the same time, the color-triplet Higgs scalars in $H_5$ also get a mass and, because they can mediate proton decay, this has to be large enough in order not to clash with the proton life-time \cite{Ellis:1978xg}. This mass-scale separation between the Higgs doublet and the color-triplet Higgs, that sit in the same multiplet, gives rise to the well-known triplet-doublet-splitting problem in GUTs \cite{Mohapatra:1997sp,Dimopoulos:1981zb}. Many possible solutions to address the issue have been conceived in the context of supersymmetric extensions of GUTs~\cite{Dimopoulos:1981zb,Witten:1981kv,Georgi:1981vf,Nanopoulos:1982wk,Dimopoulos:1982af,Masiero:1982fe,Inoue:1985cw,Barr:1997pt,Witten:2001bf,Kawamura:2000ev,Hall:2001pg}. 
In our construction of CGUT in (\ref{CFTSU(5)}) and the corresponding potential (\ref{potCFTSU5}), we assumed that the interaction between $H_5$ and $\Sigma$ are negligible and does not play a role. This means there is no relevant contribution to the mass terms of the color-triplet Higgs and SM Higgs doublet form the VEV of the adjoint field $\Sigma$. Instead, the Higgs doublet mass scale will be generated through the scalar of the hidden sector (see the potential (\ref{bosf}) and section~\ref{sec:Higgs_ele}). As far as inflation is concerned, all what we need is a coupling between $\Sigma$ and the inflaton, which generates a CW potential for the inflaton. We leave the study of a more general potential with $\Sigma$-$H_5$ and $\phi$-$H_5$ interactions fully included, together with the generalization to a supersymmetric framework, for future research on the subject. However, we explore a possible connection between CGUT and the doublet-triplet splitting problem in appendix~\ref{sec:doublet_triplet}.

Finally, our present framework of inflation  can be straightforwardly extended to $\text{SO(10)}$ by promoting the fields $\LF \phi,\,\chi \RF$ as singlets of $\text{SO(10)}$ and conformally coupling them to the 45-plet adjoint field. The symmetry breaking patterns in this context are rich~\cite{Croon:2019kpe}, for example 
within the two-stage pattern $\text{SO(10)}\to\text{G}_{422}=\text{SU}(4)_{c}\times\text{SU}\left(2\right)_{L}\times\text{SU}\left(2\right)_{R}\to\textrm{SU}(5)\to\textrm{SU}(3)_{c}\times\textrm{SU}(2)_{L}\times\textrm{U}(1)_{Y}$, there will be more observables associated with the intermediate group such as 
 generation of primordial
monopoles \cite{Senoguz:2015lba}. 
\section{Inflaton interactions with the dark sector}
\label{sec:Infl_matter}
In this section we introduce the interactions between the inflaton and the dark sector in order to describe post-inflationary production of the bulk of the matter content in the universe. We consider the dark species to be both fermions and scalars. Since the non-vanishing VEV of the inflaton generates a mass term for each particle coupled to it, we avoid SM particles to have such an interaction. Indeed, we aim at keeping the SM particles mass generation as still driven by the Higgs mechanism, namely a potential of the form (\ref{higgs_0_pot}). However, since we start with a conformal theory, a mass scale for the Higgs boson has to be generated, together with a way to produce the SM degrees of freedom in the post-inflationary phase. To this end, we allow a portal coupling consistent with conformal symmetry between the dark scalar and the SM Higgs,  so that the energy stored in the inflaton can partly leak into the SM model and produce the corresponding reheating. In so doing, we shall see that an intermediate energy scale between the inflaton mass and the electroweak scale is found. The dark scalar shall completely decay into Higgs bosons, whereas the rest of the inflaton energy is responsible for the dark matter relic density made of stable dark fermions. 

The explicit form of the interactions are comprised in the
the second term of the conformal potential in eq.~(\ref{pot_infl_reheat})\footnote{We assume the kinetic terms and non-minimal couplings of matter sector to be negligible, hence we do not write them in the action (\ref{CFTSU(5)}). }, which reads
\begin{equation}
\begin{aligned}
V(\phi,\hbox{\scriptsize SM},\hbox{\scriptsize DS}) \equiv &\, V (\phi,\chi,\psi,S,H)\\= &\, Y_\psi f_{\psi}\left(\frac{\phi}{\chi}\right) \phi \, \bar{\psi} \psi\, -\lambda_{S1}f_{S1}\left(\frac{\phi}{\chi}\right)\phi^2 S^{\dagger}S+\lambda_{S2}f_{S2}\left(\frac{\phi}{\chi}\right) \LF S^\dagger S \RF^2  \\ 
& - \lambda_{HS}\LF S^\dagger S \RF\LF H^\dagger H \RF + \lambda_H \LF H^\dagger H \RF^2\,,
\label{bosf}
\end{aligned}
\end{equation}
where $\psi$ and $S$ are the dark fermion and dark scalar fields, $H$ is the SM Higgs boson doublet, $\lambda_{HS}$ is the portal coupling between the Higgs boson and the dark scalar, $\lambda_H$ is the Higgs self-interaction.\footnote{We take the dark sector particles to be Dirac fermions and complex scalars. This way we allow for a charge in the dark sector, even though we do not exploit or elaborate the details of such a feature in this work.}  
A Yukawa coupling of the form $\bar{\psi} S \psi$ is allowed by the conformal symmetry, however we do not consider it here. We adopt this choice in order to keep a simple implementation and not to introduce an additional parameter in the model. Moreover, we choose not to introduce a conformal function of the form $f(\phi / \chi)$ in the terms that comprise the SM field $H$, having in mind a closer connection between the inflaton and the dark sector. 

The  conformal symmetry of the whole action (\ref{CFTSU(5)}) with the potential (\ref{bosf}) can be preserved by implementing the following transformations on the matter fields
\begin{equation}
S\to\Omega^{-1}(x) S\,,\quad H\to \Omega^{-1}(x)H \, , \quad \psi \to \Omega^{-3/2}(x) \psi \,.
\label{newcon}
\end{equation}
The functions $f_\psi, \, f_{S1},\,f_{S2} $ are conformally invariant with respect to transformation of the fields $\phi,\,\chi$ and we take them of a similar form as in eq.~(\ref{fPchi})
\begin{equation}
\begin{aligned}
\quad f_\psi=\LF 1-\frac{\phi^2}{\chi^2}  \RF^{\alpha} \, ,  \quad f_{S1} = \LF 1-\frac{\phi^2}{\chi^2} \RF^{\beta_1},\,\quad f_{S2} = \LF 1-\frac{\phi^2}{\chi^2}  \RF^{\beta_2} \, .
\label{conf_coupl_matter}
\end{aligned}
\end{equation}
There is however a relevant difference given by the exponents $\alpha$, $\beta_1$ and $\beta_2$, that we take to be positive. Such a choice of the conformal couplings allows for suppressing the couplings between the inflaton and the dark particles with respect to inflaton-GUT fields couplings during inflation. This is important  in order to let the GUT fields to be the relevant source for the effective potential of the inflaton during the slow-roll dynamics. As anticipated in section~\ref{sec:conf_infl}, we require $V\LF \phi,\,\chi,\,\Sigma \RF$ in the tree level potential (\ref{pot_infl_reheat}) to be dominant term during inflation. This can be achieved by imposing $\alpha,\,\beta_1 \gg 2$ and $\beta_2 \gg 2$ in the couplings (\ref{conf_coupl_matter}). Indeed, as we learned in section~\ref{sec:conf_infl}, inflation happens when $\phi \to \sqrt{6}M$, which automatically implies the conformal couplings in (\ref{conf_coupl_matter}) to be suppressed during the period of inflation. As we shall see, having $\alpha,\,\beta_1 \gg 2$ also give dark particle masses that are (much) smaller than the inflaton mass, see section~\ref{sec:dark_masses}, whereas $\beta_2 \gg 2$ ensures that the effective self-coupling of $S$ scalars is suppressed in comparison with the self-coupling of the $\Sigma$ field. In order to make the exponents $\alpha, \beta_1, \beta_2$ fully responsible for the strength of the couplings and reduce the number of parameters, we fix the couplings $Y_\psi$, $\lambda_{S1}$ and $\lambda_{S2}$ in (\ref{bosf}) to unity. 

\subsection{Dark particles masses and tree-level interactions}
\label{sec:dark_masses}
The functions $f_\psi$, $f_{S1}$ and $f_{S2}$ in the potential (\ref{bosf}) lead to a dynamical generation of the masses of the dark particles  when the inflaton acquires its VEV, together with tree-level couplings that induce the inflaton decays into dark fermion/scalar pairs.  To the best of our knowledge, the conformal couplings  (\ref{conf_coupl_matter}) have not been exploited to provide the features of the dark particles.  

In the following, we assume small perturbations around the inflaton VEV,
$\phi =  v_\phi +  \delta \phi + \cdots \, = v_\phi + \Phi + \cdots$,
where we define $\delta \phi \equiv \Phi$ to be the dynamical inflaton field. With the aim of determining the dark particle masses and the tree-level interactions for the inflaton decays, we single out the constant and linear terms in $\Phi$ from the conformal couplings in eq.~(\ref{conf_coupl_matter})
\begin{eqnarray}
&&f_{\psi} = \gamma^{2 \alpha} - 2 \alpha \frac{\Phi}{\sqrt{6}M} \gamma^{2(\alpha-1)} \left( 1-  \gamma^2 \right)^{\frac{1}{2}}  \, ,
\end{eqnarray}
for the fermion dark matter $\psi$.
The same holds for the scalar particle couplings when substituting with the corresponding coefficients $\beta_1$ and $\beta_2$, however we consider only the leading $\Phi$-independent term for $\beta_2$ in the following. 
Next, we expand the inflaton field around $v_\phi$ in the vertices involving the fermionic and bosonic matter in eq.~(\ref{bosf}), so that one finds the dark matter fermion and scalar mass to be
\begin{eqnarray}
m_\psi = \, \gamma^{2 \alpha} v_\phi \, , \quad \mu_S^2 =      \gamma^{2\beta_1} v_\phi^2 \, .
\label{m_ferm_bos}
\end{eqnarray}
Let us remark that, at this first stage, we can single out the symmetric potential for dark scalar from (\ref{pot_infl_reheat}) that reads\footnote{If one assumes there is no gauge symmetry and corresponding charge carried by the $S$ field, then a $Z_2$ symmetry is the one to be broken in the potential (\ref{pot_B_0}). }
\begin{equation}
V_S = -\mu_S^2(S^\dagger S)+\lambda_{S} \LF S^\dagger S\RF^2\,.
\label{pot_B_0}
\end{equation}
where $\mu_S^2 =  v_\phi^2 \gamma^{2\beta_1} >0$ and 
$\lambda_{S} \equiv \gamma^{2\beta_2}$ and the  potential for $S$ shows up in the form of (\ref{higgs_0_pot}) with an overall negative mass term.
Correspondingly, the potential is destabilised and  the $S$ field acquires a VEV, namely $v^2_S =\mu_S^2/ \lambda_S$, and the physical mass of the dark scalar reads $m_S=\sqrt{2}\mu_S$.
   
\begin{figure}[t!]
\centering
\includegraphics[scale=0.57]{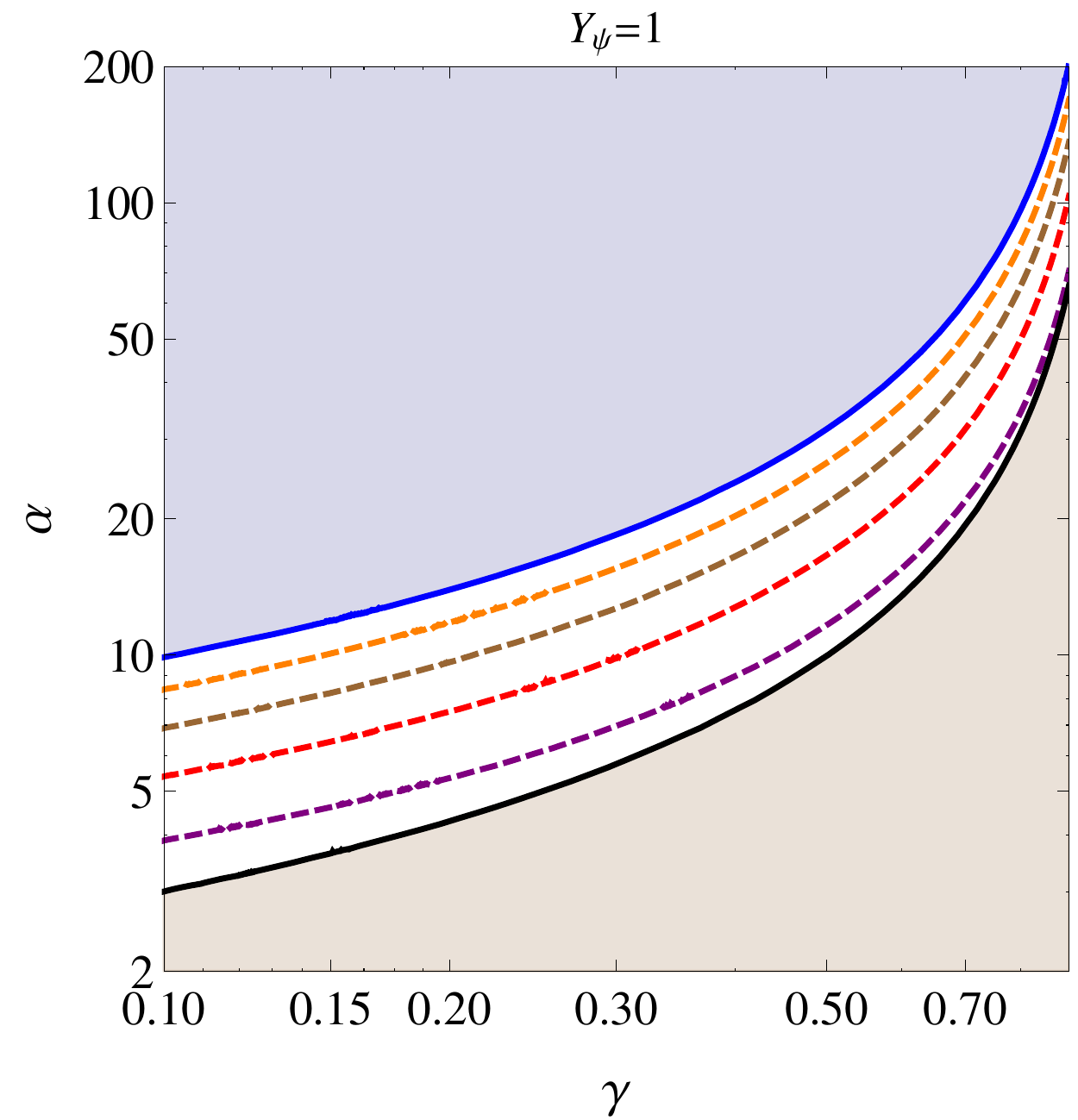}
\hspace{0.3 cm}
\includegraphics[scale=0.572]{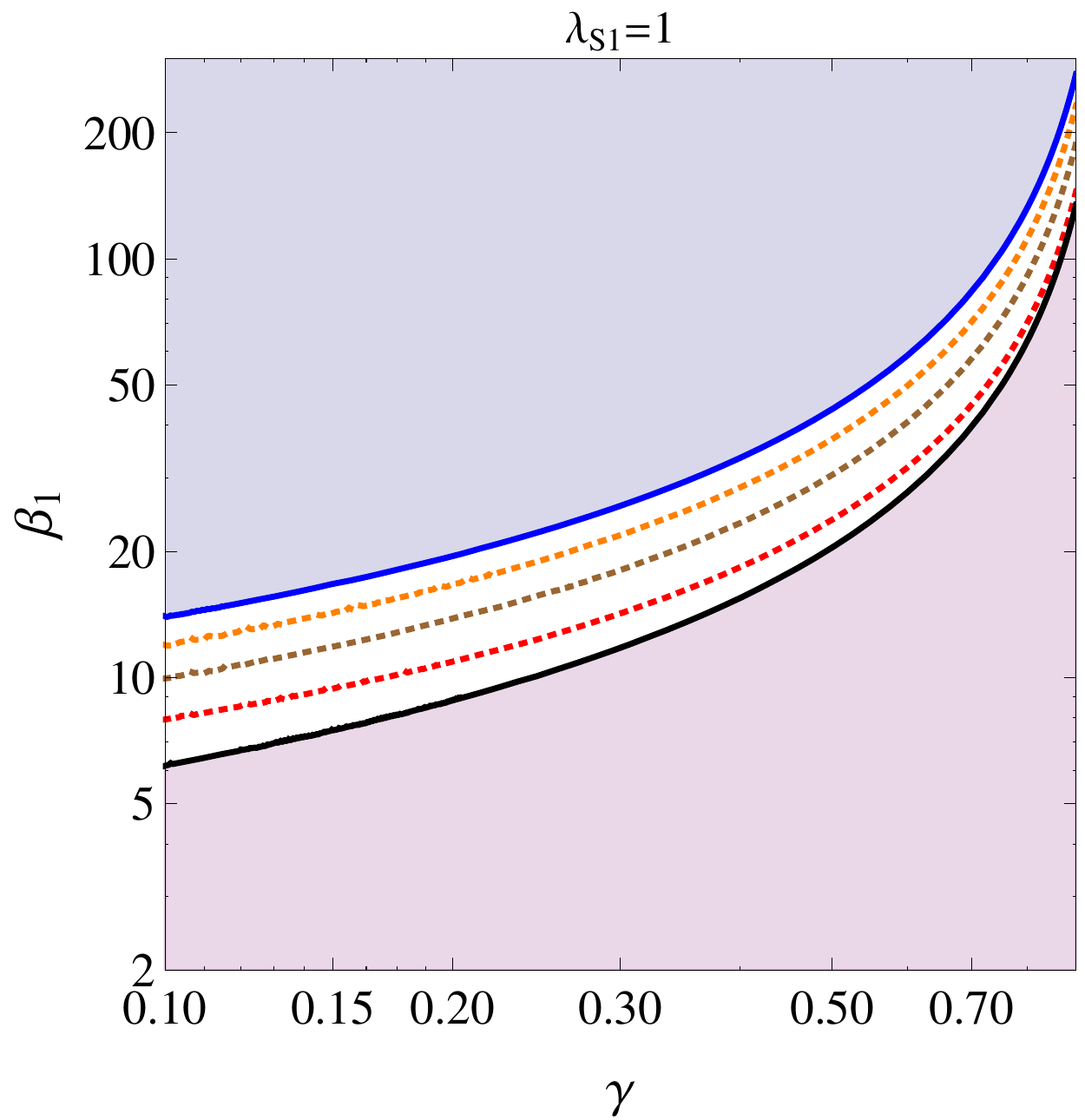}
\caption{\label{fig:figMass}The masses for fermion and boson particles as given in eqs.~(\ref{m_ferm_bos}) with $m_S = \sqrt{2}\mu_S$. The black solid lines stand for $2 m_S=M_\Phi$ and $2 m_\psi=M_\Phi$, and  the shaded areas account for $2 m_S>M_\Phi$ and $2 m_\psi>M_\Phi$. The solid blue lines and the corresponding shaded area implements the BBN constraint. From top to bottom, the dashed lines stand for $m_\psi=10^3, \, 10^6, \, 10^9, \, 10^{12}$ GeV, whereas the dotted lines correspond to $m_S=10^8, \, 10^{10}, \, 10^{12}$ GeV.}
\end{figure}

For this implementation to work, we assume that the dark scalar does not get 1-loop CW quantum corrections due to its coupling with the inflaton field. Therefore, we consider the self-coupling of the dark scalar to be larger than its coupling to inflaton field,
   that can be easily realized for $\beta_2 \ll \beta_1$ and  $\gamma<1$ (we anticipate that this condition holds when the observed Higgs boson mass is generated form the VEV of the dark scalar, see section~\ref{sec:Higgs_ele}). 
From eq.~(\ref{m_ferm_bos}) and $m_S=\sqrt{2}\mu_S$,  we see that the inflaton generates bosonic states which are heavier than the fermionic ones, whenever the same exponent $\alpha=\beta_1$ is inserted in the conformal couplings (\ref{conf_coupl_matter}) 
 \begin{equation}
     \frac{m_S}{m_\psi}= \frac{\sqrt{2}}{\gamma^{\alpha}} \, .
 \end{equation}
We show the dark fermion and boson masses in Fig.~\ref{fig:figMass} in the parameter space $(\gamma,\alpha)$ and $(\gamma,\beta_1)$ respectively, where the coloured dashed and dotted lines correspond to benchmark mass values for the dark particles.

Our aim is to consider the inflaton decays into particles of the hidden sector. To this end, we need to write the tree-level vertices for the processes $\Phi \to \psi \psi$ and $\Phi \to ss$, where $s$ stands for the dynamical dark scalar field  with mass $m_S=\sqrt{2}\mu_S$. From the potential in eq.~(\ref{bosf}) we obtain, by singling out the linear terms in $\Phi$,  the following interaction Lagrangian
\begin{eqnarray}
\mathcal{L}_{\phi,\hbox{\scriptsize DS}} = -    \,  \gamma^{2(\alpha-1)}\left[  \gamma^2 (2 \alpha +1)  -2 \alpha  \right] \bar{\psi} \psi \Phi \, 
- \,  v_\phi \gamma^{2(\beta_1-1)} \left[ \gamma^2 (\beta_1 +1)  - \beta_1  \right] s^2  \, \Phi \, , 
\label{int_infl_b}
\end{eqnarray}
where we remind that the physical scalar is the one obtained after the symmetry breaking with $S=(v_S+s)/\sqrt{2}$ in unitary gauge.
In order to make the notation simpler, we define the couplings 
\begin{equation}
    y_\psi \equiv \gamma^{2(\alpha-1)}\left[  \gamma^2 (2 \alpha +1)  -2 \alpha  \right] \, , \quad y_S =  \frac{v_\phi}{M_{\textrm{P}}}  \gamma^{2(\beta_1-1)} \left[ \gamma^2 (\beta_1 +1)  - \beta_1  \right] \, ,
    \label{int_infl_c}
\end{equation}
and we factor out the Planck mass from the scalar coupling so to have a dimensionless quantity.
In order for the inflaton to decay perturbatively into the dark sector, two conditions have to be met. First, the kinematic conditions $M_\Phi > 2 m_\psi$ and $M_\Phi > 2 m_S$ has to hold.\footnote{In our study we consider the coupling $y_S< 36\pi^2 M_{\Phi}^2/M_{\textrm P}^2$ in which case the effects of parametric resonance are negligible \cite{Dolgov:1989us,Traschen:1990sw,Kofman:1994rk}.}  This imposes a condition on $\alpha$ and $\beta_1$ for different values of $\gamma$ in the range $0.1 \leq \gamma \leq 0.9$. Both in the right and left plot in Fig.~\ref{fig:figMass}, we indicate the condition $2 m_S =M_\Phi$ and $2 m_\psi =M_\Phi$ with a black solid line. The shaded region below the solid black line is then not relevant to our study. Let us stress that the kinematic condition for the inflaton decays to happen points to values of $\alpha$ and $\beta_1$ that automatically suppress the relevance of the inflaton-dark sector coupling during inflation as desired.  The decay width of the inflation into fermion and boson pairs,  $\Gamma_{\Phi\to \psi \psi}$ and $\Gamma_{\Phi\to ss}$, read respectively
\begin{eqnarray}
    \Gamma_{\Phi\to \psi \psi} &=& \frac{M_\Phi y_\psi^2}{8 \pi} \left( 1-\frac{4m_\psi^2}{M^2_\Phi}\right)^{3/2} = \frac{M_\Phi \alpha_\psi}{2 } \left( 1-\frac{4m_\psi^2}{M^2_\Phi}\right)^{3/2} \, ,
    \label{decay_phi_to_psi}
    \\
    \Gamma_{\Phi\to ss} &=& \frac{M_{\textrm{P}}^2 y_S^2}{ 16 \pi M_\Phi} \left( 1-\frac{4m_S^2}{M^2_\Phi}\right)^{1/2} = \frac{M_{\textrm{P}}^2 \alpha_S}{4 \, M_\Phi } \left( 1-\frac{4m_S^2}{M^2_\Phi}\right)^{1/2} \, .
    \label{decay_phi to_b}
\end{eqnarray}
where $\alpha_S \equiv y_S^2 /(4 \pi)$ and $\alpha_\psi \equiv y_\psi^2/(4 \pi)$, and
the couplings $y_\psi$ and $y_S$ have been given in eq.~(\ref{int_infl_c}). The second condition for perturbative decay, $\alpha_\psi \, , \alpha_S < 1$ is well met in all the parameter space of the model and does not pose any constraint.

There is an additional important aspect that imposes a condition on the model parameters through the inflaton decay widths in eqs.~(\ref{decay_phi_to_psi}) and (\ref{decay_phi to_b}). This is the constraint given by the BBN nucleosynthesis time scale. Indeed, any heavy particle in the early universe has to decay before this epoch and ever since the standard thermal history is expected to hold. The decays of the inflaton into the hidden sector has to satisfy this condition as well because the SM plasma is in turn generated  from the decays of  $S$ scalars though the portal coupling $\lambda_{HS}$ in this model. Therefore, we require $\tau_{\Phi \to ss} \, , \tau_{\Phi \to \psi \psi}   < 0.1 
  $ s, 
  and accounting for the conversion 1 GeV$ \simeq 1.52 \times 10^{24} s^{-1}$ we find \begin{equation}
  \Gamma_{\Phi\to \psi\psi} \, , \; \Gamma_{\Phi\to ss}   > 6.6 \times 10^{-24} \; \hbox{GeV} \, .
  \label{BBN_infl}
  \end{equation}
In Fig.~\ref{fig:figMass}, we show the BBN constraint with a blue solid line and corresponding blue shaded region, where the model provides inflaton decay widths not compatible with (\ref{BBN_infl}). In summary, the model remains viable in the not-shaded regions. The dashed lines corresponds to fixed-mass hypotheses for the dark particles, and we provide them for orientation. As far as the dark fermion is concerned, the dashed lines stand for $m_\psi=10^3, \, 10^6, \, 10^9, \, 10^{12}$ GeV, from top to bottom.  Masses down to 1 GeV are viable for the dark fermions that correspond to the boundary given by the solid blue line from BBN.\footnote{Such mass range for the stable dark fermions implies that the bound on  $N_{\hbox{\scriptsize eff}}$ is not affected because these particle are heavy and cannot qualify as relativistic degrees of freedom at the BBN epoch.}
The situation for the dark scalar is different due to the dependence of the decay width in eq.~(\ref{decay_phi to_b}) and the mass $m_S$ on $\gamma$ and $\beta_1$.  In the right plot of Fig.~\ref{fig:figMass}, the solid blue line corresponds to $m_S=10^6$ GeV making the allowed dark scalar masses  much heavier than the fermion case. From top to bottom, the dotted lines correspond to $m_S=10^8, \, 10^{10}, \, 10^{12}$ GeV. 

\subsection{Higgs mass and electroweak symmetry breaking from the dark sector}
\label{sec:Higgs_ele}
Starting with the scale invariant potential (\ref{pot_infl_reheat}), the inflaton VEV is responsible for the generation of the dark scalar mass term $\mu_S$, which in turn generates a Higgs mass term $\mu_H$ through the portal coupling $\lambda_{HS}$. Since we assumed a negligible coupling between the adjoint field $\Sigma$ and the fundamental $H_5$ in our CGUT framework, the generation of the Higgs mass term is entirely due to the dark scalar. Our aim is to further  constrain  the  model  parameter  space  with the  condition  given by  the  observed Higgs boson mass $m_H=125.18 \pm 0.16$ GeV \cite{Tanabashi:2018oca}. Similarly to our assumption for the dark scalar in section~\ref{sec:dark_masses}, we assume the Higgs portal coupling to the $S$ field to be much smaller than the self-coupling $\lambda_H$ so to protect the Higgs field to get substantial CW corrections.\footnote{At tree level and at the electroweak scale $\lambda_H \approx m_H^2 g_2^2/(8 m_W^2) \approx 0.13$, where $m_W$ is the W boson mass and $g_2$ is the gauge coupling of SU(2) SM gauge group.} As shown below, this assumption is going to be well satisfied since the portal coupling $\lambda_{HS}$ turn out to be very small.

Starting from the potential (\ref{bosf}) and after $\phi$ and $S$ symmetry breaking,  we can consider the $s$-independent terms and write the following Higgs potential in the form of eq.~(\ref{higgs_0_pot})
\begin{equation}
\begin{aligned}
    V_H &=-\lambda_{HS} \frac{v_S^2}{2} (H^\dagger H) + \lambda_H (H^\dagger H)^2 =-\mu_H^2 (H^\dagger H) + \lambda_H (H^\dagger H)^2 \, ,
    \end{aligned}
    \end{equation}
  where in the second step we define $\mu_H^2 \equiv \lambda_{HS}v_S^2/2$. This way a negative mass term  appears in the Higgs potential and the Higgs boson undergoes the spontaneous breaking of the electroweak symmetry. The physical Higgs mass reads  $m_H=\sqrt{2}\mu_H$, and in terms of $v_S$ and $v_\phi$ we obtain 
    \begin{equation}
        m_H^2=\lambda_{HS} v_S^2= \lambda_{HS} \, \gamma^{2(\beta_1-\beta_2)} v_\phi^2 \, ,
        \label{Higgs_mass_eq}
    \end{equation}
   Finally, by using eq.~(\ref{phiVev}), we can write the Higgs mass in terms of the additonal model parameters $\gamma$, $\beta_1$ and $\beta_2$   \begin{equation}
        m_H^2= 
          6 \lambda_{HS} \, \gamma^{2(\beta_1-\beta_2-1)} (1-\gamma^2)m_{\hbox{\tiny P}}^2 .
        \label{higgsms}
    \end{equation}
With the expression (\ref{higgsms}) we can now explore the model parameter space compatible with the observed Higgs mass. The four parameters $\gamma$, $\beta_1$, $\beta_2$ and $\lambda_{HS}$ are involved. In general the Higgs boson and the dark scalar shall mix. However, due to the quite large masses for the dark scalar $m_S>10^6$ GeV considered in this work (see discussion in section~\ref{sec:dark_masses}), we can assume a small mixing. This approximation allows us take the observed Higgs mass $m_H$ as the mass eigenstate of the light scalar in the two-scalar mixing, and use eq.~(\ref{higgsms}) as a constraint. 
We give some details on the scalar mixing in Appendix~\ref{sec:scalar_mix}. 

\begin{figure}[t!]
\centering
\includegraphics[scale=0.78]{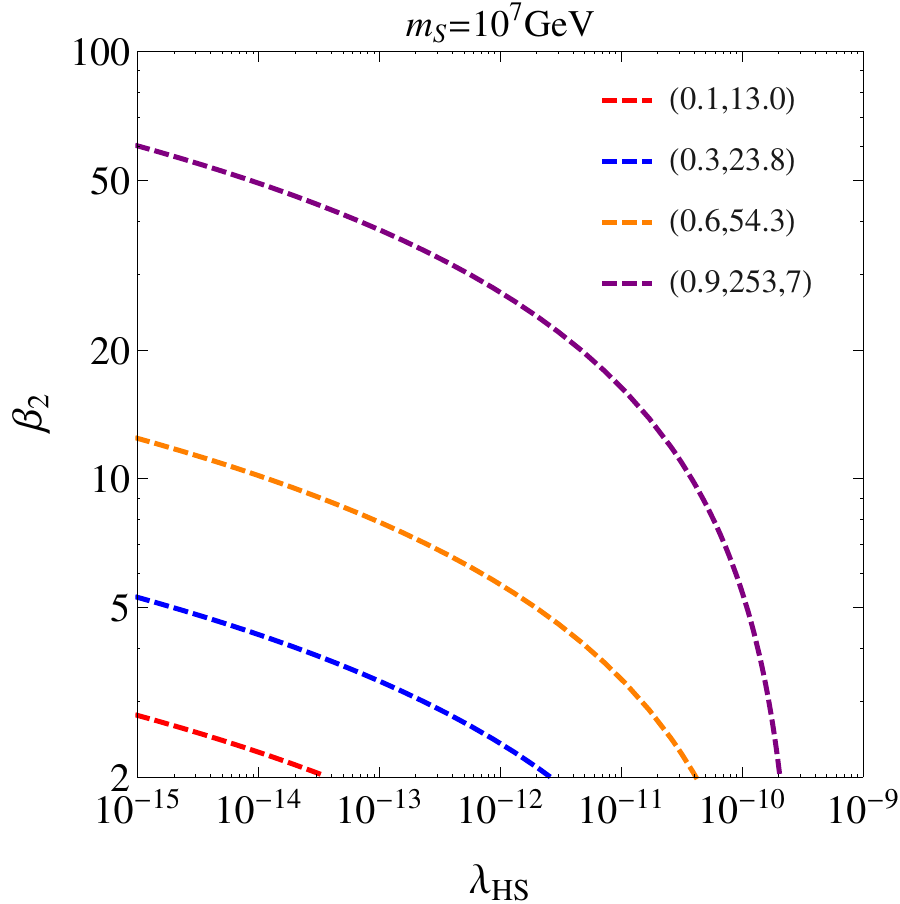}
\hspace{0.3 cm}
\includegraphics[scale=0.79]{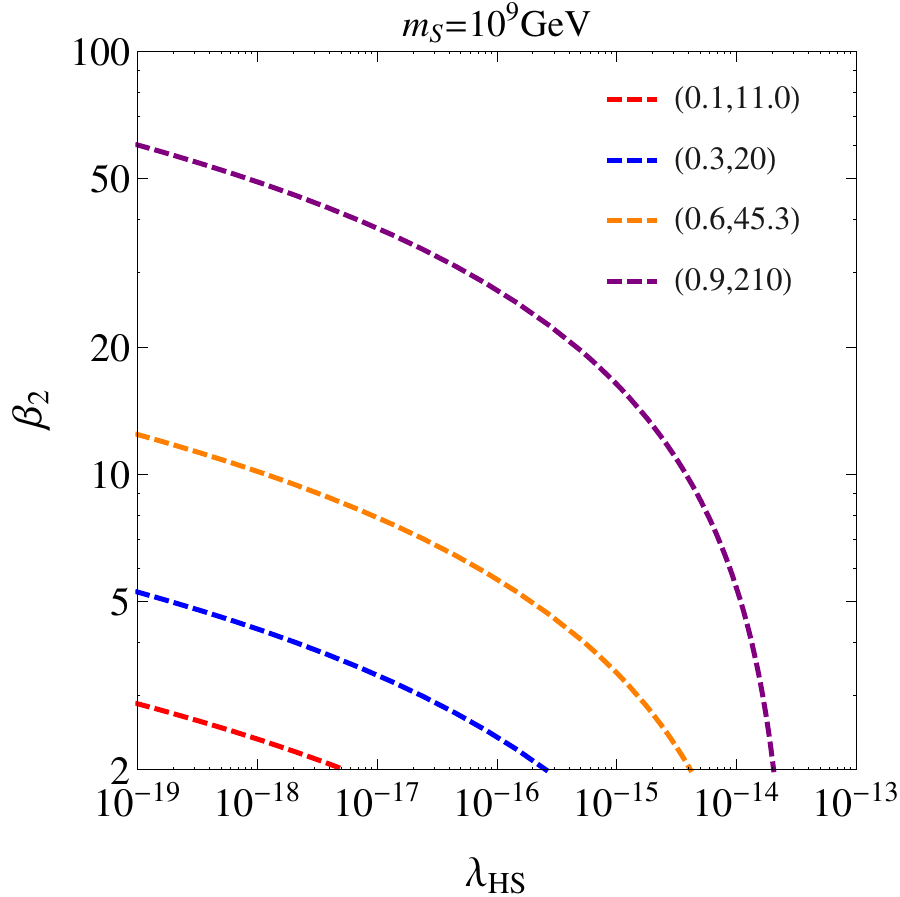}
\caption{\label{fig:figMB_Higgs} Parameter space $(\lambda_{HS},\beta_2)$ compatible with the Higgs boson mass $m_H=125.18$ GeV. Two benchmark values $m_S=10^7$ GeV and $m_S=10^9$ GeV are considered in the left and right plot respectively. The solid-dashed lines correspond to four values of $(\gamma,\beta_1)$ which are fixed to comply with the given dark scalar masses.}
\end{figure}

We orient ourselves by taking two benchmark values for the dark scalar mass $m_S=10^7$ GeV and $m_S=10^9$ GeV. As shown in the legend of Fig.~\ref{fig:figMB_Higgs}, we obtain pairs ($\gamma, \beta_1$) that comply with the given choice of the dark scalar mass. We saw already in Fig.~\ref{fig:figMass} that, for a fixed $\gamma$, the larger $\beta_1$ the smaller $m_S$.  In Fig.~\ref{fig:figMB_Higgs}, the four dashed lines show the contours in the $(\lambda_{HS},\beta_2)$ plane compatible with the Higgs mass for different values of $\gamma$. As expected, because of the large separation of the dark scalar and Higgs masses, the portal coupling is very small.
 One may see that the allowed parameter space for $\lambda_{HS}$ shifts to smaller values when going from $m_S=10^7$ GeV to $m_S=10^9$ GeV (roughly four orders of magnitude). The result can be understood as follows. Larger values of the dark scalar mass correspond to smaller values of $\beta_1$. Keeping $\beta_2$ fixed, the less effective suppression from $\gamma^{2 \beta_1 -2 \beta_2 -2}$ in (\ref{higgsms}) has to be compensated by smaller values of the portal couplings. The same reasoning can be used to understand the trend of each dashed curve in both plots in Fig.~\ref{fig:figMB_Higgs}. 
 The largest possible value for $\lambda_{HS}$ is given in the limiting case $m_S=10^6$ GeV, that we recall is a boundary situation to comply with BBN, and we find it to be $\lambda_{HS} \simeq  10^{-7}$ .
 
Finally we give the expression for the decay width of the dark scalar into a Higgs boson pair. This is a key ingredient in the analysis that we carry out in the next section as regards the SM reheating. Indeed, the inflaton is not directly coupled to the SM sector in this model. The dark scalar plays the role of a mediator that transfers some of the energy stored in the inflaton at the end of inflation into SM degrees of freedom. After the symmetry breaking of the dark sector and the electroweak symmetry, a vertex that mediates the decay $s \to hh$ is induced and the expression for the width is  
\begin{equation}
  \Gamma_{s \to hh}\equiv \Gamma_S = \frac{\lambda_{HS}}{32 \pi} \frac{m_H^2}{m_S} \sqrt{1-\frac{4 m_H^2}{m_S^2}} + \sin^2(\theta) \Gamma_h(m_S) \, ,
  \label{Higgs_width_1}
  \end{equation}
  where we have used the condition $m^2_H=\lambda_{HS} \, v_S^2$ to express the first term in eq.~(\ref{Higgs_width_1}).
  The first term in the decay width amounts to the direct decay of the dark scalar into Higgs pairs (in our model $m_S \gg 2 m_H$ and the square root can be set two unity). The latter term is triggered by the $S$-Higgs boson mixing  and we take the off-shell Higgs boson width as in refs.\cite{Cline:2013gha,Dittmaier:2011ti}. However, we checked that its contribution is negligible in all the parameter space of interest due to the small mixing angle, which reads  in the limit $\lambda_{HS} \ll \lambda_H, \lambda_S$, as follows
\begin{equation}
\theta \simeq \frac{\lambda_{HS} v_H v_S}{m_S^2-m_H^2} = \frac{\lambda_{HS}^{3/2}}{2^{\frac{3}{2}} \lambda_S } + \cdots
\label{mix_angl_0}
\end{equation}
The dots stand for corrections in $m_H^2/m_S^2 \ll 1$ and we used the relations between VEVs, masses and four-scalar couplings to obtain the last expression in eq.~(\ref{mix_angl_0}). 

An important comment is in order. We see that the conformal coupling between the inflaton and the dark scalar in (\ref{bosf}), together with the constraint from the BBN time scale, allow for massive states $m_S \simg 10^6$. Let us assume that we remove the dark scalar and couple instead the Higgs boson directly to the inflaton field with a conformal coupling. Then, it seems not possible to generate a scalar with a mass of order $10^2$ GeV (as the SM Higgs boson is) and be consistent with BBN at the same time. This is manly due to the large inflaton VEV that has to be compensated with very large exponents in the conformal couplings to generate small scalar masses. However, large exponents bring in turn to small tree-level couplings between the inflaton and the Higgs boson, and eventually induce a too small decay width not compatible with the BBN time scale.

\begin{figure}[t!]
    \centering
   \includegraphics[scale=0.595]{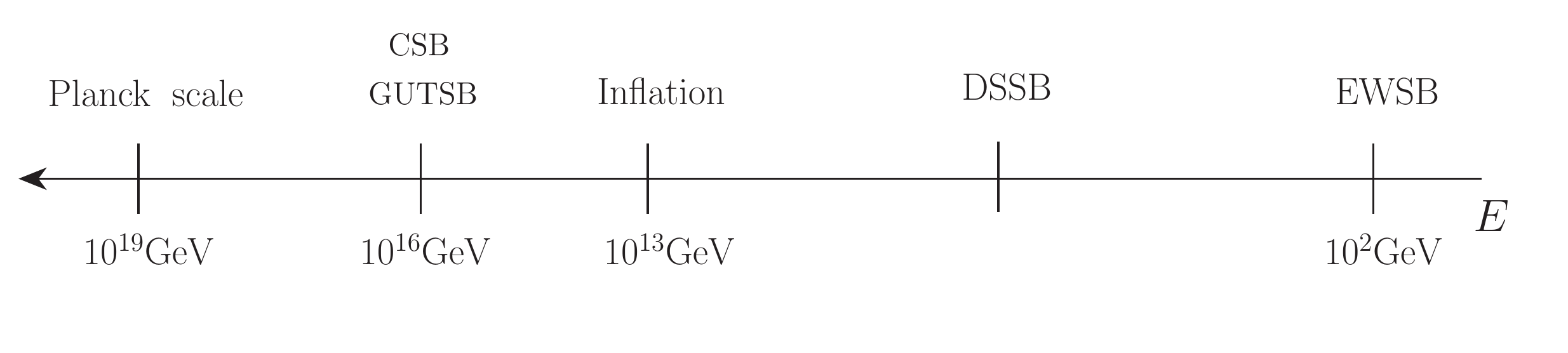}
    \caption{In this plot we depict the associated hierarchy of energy scales (from left to right) and symmetry breaking patterns in our model. We obtain Starobinsky-like inflation after Conformal Symmetry Breaking (CSB) and GUT Symmetry Breaking (GUTSB) respectively. Later on, a Dark Sector Symmetry Breaking (DSSB) occurs at some intermediate scale between the inflation scale and Electroweak Symmetry Breaking (EWSB) scale.}
    \label{fig:Esscales}
\end{figure}

 Having delineated the main features and compelling parameter space for the model, we summarize the energy scales and symmetry breaking patterns discussed so far in Fig.~\ref{fig:Esscales} . We shall address the thermal history after CGUT inflation in the next  section~\ref{sec:numerical_final}. 
\section{Dark matter relic density and SM reheating}
\label{sec:numerical_final}
In this section we address the production of dark particles, both the dark fermions and dark scalars directly coupled to $\phi$, and the SM degrees of freedom after inflation. The picture is the following. At the end of inflation, the universe has been given the initial kick for its expansion, however it is left empty and without any particle but $\phi$.  The energy stored in the inflaton field is then converted into other particles through its decays, and an equilibrium heat bath of SM particles is formed.   This epoch is called reheating and the inflaton field decays into SM particles can happen either  directly  or  via  mediator  fields~\cite{Kofman:1994rk,Kofman:1997yn}.  In order to avoid spoiling predictions of the standard BBN, the SM has to become the dominant energy density component before temperatures as low as 4 MeV~\cite{Kawasaki:2000en,Hannestad:2004px,Ichikawa:2005vw,DEBERNARDIS2008192}.

In this work, we prevent a direct coupling between $\phi$ and SM particles. The inflaton only decays into  the hidden sector and a population of dark fermions and scalars is induced. The fermions are inert and stable and make up for the present-day dark matter relic density $\Omega_{\hbox{\tiny DM}}h^2 = 0.1200 \pm 0.0012$ \cite{Aghanim:2018eyx}. However, we can still generate a SM thermal bath via the portal coupling $\lambda_{HS}$ between the dark scalar and the Higgs boson. The heavy dark scalars start to decay as soon as they are produced by the inflaton, and they work as a mediator for transferring energy from the inflaton to the SM. The Higgs boson is understood to generate the SM relativistic degrees of freedom, therefore we connect the SM temperature with the Higgs boson density as implemented for example in refs.~\cite{Chu:2011be,Tenkanen:2016jic}. 

\begin{figure}[t!]
    \centering
    \includegraphics[scale=0.79]{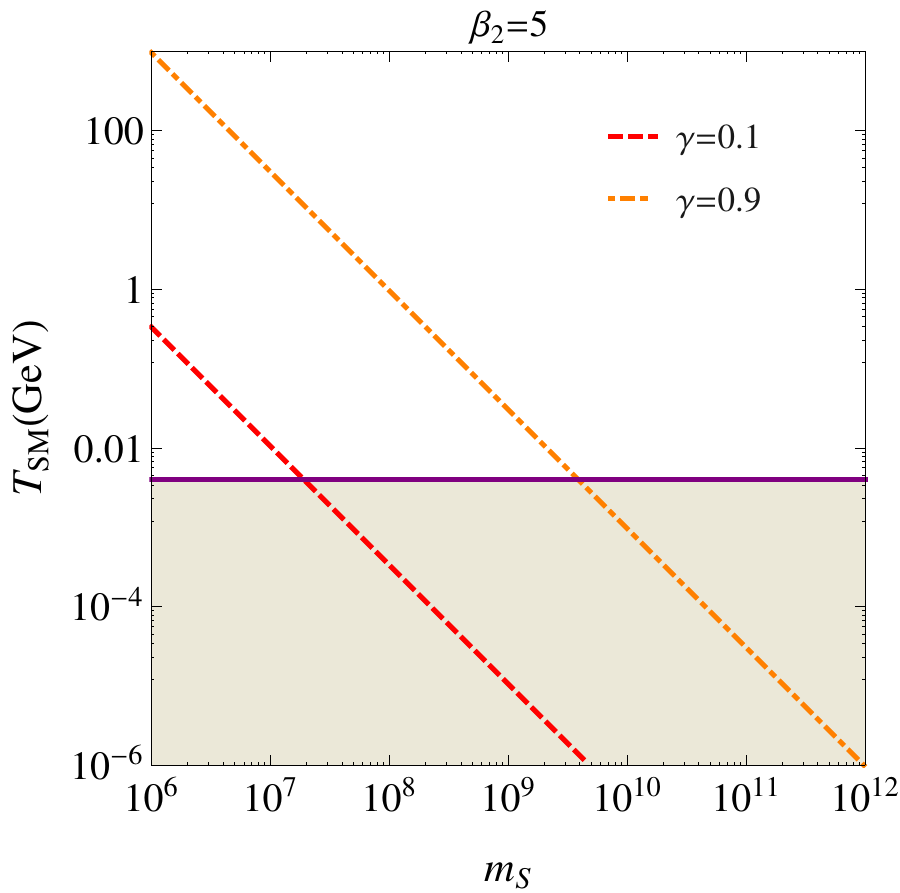}
    \hspace{0.25 cm}
    \includegraphics[scale=0.79]{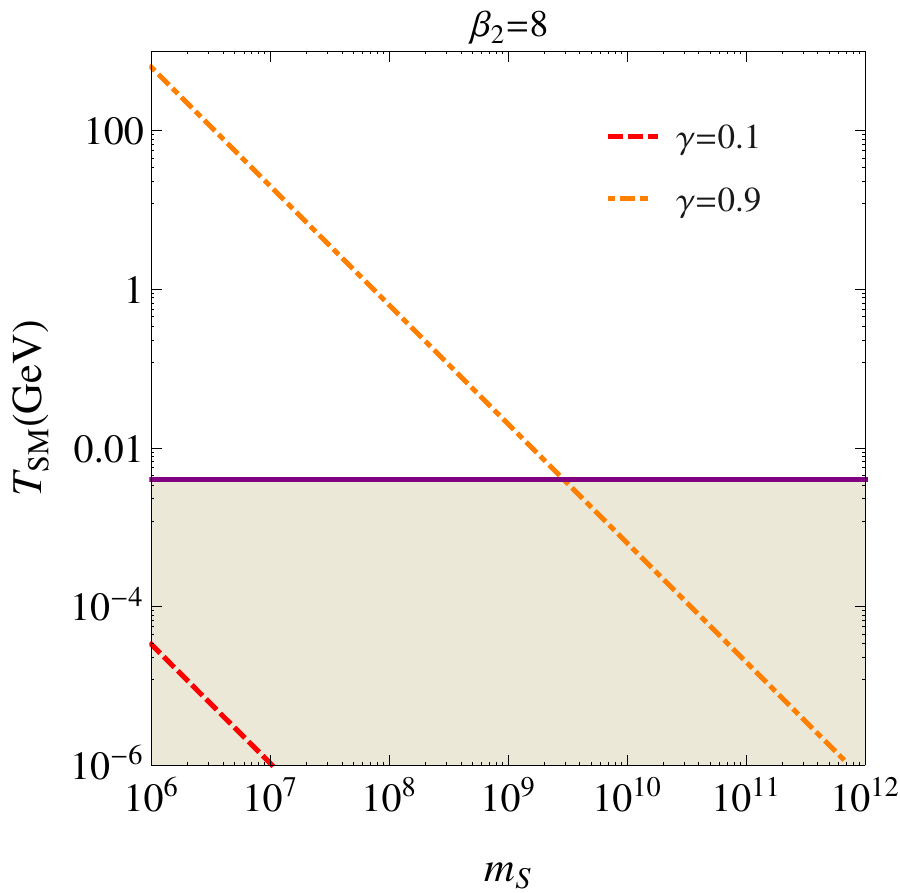}
    \caption{The instantaneous reheating temperature (\ref{SM_T_rh}) is given as function of the dark scalar mass $m_S$ for $\gamma=0.1$ and $\gamma=0.9$, red-dashed and orange dot-dashed lines respectively. In the left (right) plot $\beta_2=5$ $(\beta_2=8)$. The shaded area gives temperatures smaller than 4 MeV.}
    \label{fig:T_RH_inst}
\end{figure}
Despite we shall solve a network of Boltzmann equations to track the evolution of the dark matter and the SM temperature in section~\ref{sec:Boltzmann_numerics}, it is useful to give an estimation of the reheating temperature for the SM using the approximation of instantaneous reheating. In short, such a version of the reheating temperature can be obtained by assuming that all the available energy stored in the decaying particle is instantaneously converted into radiation.  As the SM is generated through the decays of the dark scalars, the corresponding reheating temperature reads
\begin{equation}
    T^{\hbox{\tiny RH},i}_{\hbox{\tiny SM}} = \sqrt{\Gamma_{S} M_{\textrm{P}}} \left(\frac{90}{ \pi^2 g_{\hbox{\scriptsize eff}}(T^{\hbox{\tiny RH},i}_{\hbox{\tiny SM}})} \right)^{1/4} \, ,
    \label{SM_T_rh}
\end{equation}
where the relevant decay width is given in eq.~(\ref{Higgs_width_1}) and $g_{\hbox{\scriptsize eff}}$ amounts to the number of relativistic degrees of freedom at the reheating temperature. In Fig.~\ref{fig:T_RH_inst}, we show the reheating temperature as provided by eq.~(\ref{SM_T_rh}) for the two benchmark values $\gamma=0.1$ and $\gamma=0.9$, respectively the red-dashed and dot-dashed orange lines. In the left plot, we see that there is a rather limited parameter space available to comply with  $T^{\hbox{\tiny RH}}_{\hbox{\tiny SM}} \geq 4$ MeV for $\gamma=0.1$, whereas the situation sensibly improves for larger values of $\gamma$. Also, increasing from $\beta_2=5$ to $\beta_2=8$ makes the $\gamma=0.1$ choice incompatible with a large enough reheating temperature.  The trend can be understood again by looking at Fig.~\ref{fig:figMB_Higgs}.  There we see that, for a given $m_S$, smaller values of $\gamma$ are related to smaller portal couplings $\lambda_{HS}$. Accordingly the decay width $\Gamma_S$ entering (\ref{SM_T_rh}) is smaller and so is the corresponding reheating temperature. 

In this work we assume that the dark particles do not form a thermal bath in the hidden sector. The temperature of such sector system would be measured by the radiation energy density of the dark sector. To this end, one needs thermalized relativistic degrees of freedom. However, the dark fermions are an inert component and do not interact with anything else after they are produced from the inflaton decays. Only the dark scalars could be candidates for a thermal bath in the hidden sector.  We checked that the would-be instantaneous reheating temperature, as estimated from (\ref{SM_T_rh}) with $\Gamma_{\Phi \to ss}$ instead of $\Gamma_S$, is order of magnitudes smaller than $m_S$ in the whole parameter space of interest. Hence, the dark scalar cannot qualify as thermalized relativistic degrees of freedom and we do not follow the temperature evolution in the hidden sector.\footnote{The reheating temperature is not the largest temperature obtained during the inflaton decay~\cite{Chung:1998rq,Giudice:2000ex}. The maximum temperature peaks at very early stages and then quickly decreases. There could be a limited time range in the very early stages of $\phi$ decays where the dark scalars could behave as a thermalized plasma of relativistic particles in our model. We make an approximation when neglecting this stage.} Also, we are interested in keeping  $\beta_2$ large enough to suppress the  self-coupling of $S$ during inflation, and this helps in suppressing the processes that would induce a thermalization in the dark scalar sector. A complementary framework where dark scalars form a thermal bath in the hidden sector with its own temperature is described for example in ref.~\cite{Tenkanen:2016jic}.
\subsection{Boltzmann equations}
\label{sec:Boltzmann_numerics}
In this section we solve a network of Boltzmann equations to track the abundance of the inflaton, dark fermions and dark scalars and the SM. Our main scope is to provide benchmark values in the model parameter space that reproduce both the observed relic density $\Omega_{\hbox{\tiny DM}}h^2 = 0.1200 \pm 0.0012$~\cite{Aghanim:2018eyx} and $T^{\hbox{\tiny RH}}_{\hbox{\tiny SM}} \geq 4$ MeV. The dark matter fermions are stable and are responsible for the present-day relic density in this model. They are produced non-thermally from the inflaton decays and are inert ever since. The dark scalars are produced also non-thermally from the inflaton decays. However, they in turn decay into SM Higgs bosons that we take as source of the SM component radiation (see  ref.~\cite{Tenkanen:2016jic} for a similar implementation for the generation of the SM plasma). 
 
 In the remaining of the paper, we shall not refer to an instantaneous reheating temperature as given (\ref{SM_T_rh}), rather we obtain the temperature evolution from the Boltzmann equations together with the extraction of the relic abundance of the dark fermions $\psi$. Indeed, as discussed and shown in refs.~\cite{Chung:1998rq,Giudice:2000ex,Arcadi:2011ev}, the reheating temperature  $T^{\hbox{\tiny RH}}_{\hbox{\tiny SM}}$ can be properly defined when the radiation component finally scales as $T_{\hbox{\tiny SM}} \propto a^{-1}$, where $a$ is the scale factor, whereas the temperature exhibits different power laws in earlier stages. 
The Boltzmann equations for the various components read~\cite{Chung:1998rq,Giudice:2000ex,Arcadi:2011ev}
\begin{eqnarray}
&& \dot{\rho}_\Phi + 3 H \rho_\Phi = - \Gamma_{\Phi} \rho_\Phi
\label{BE_phi}
\\
&&\dot{n}_\psi + 3 H n_\psi = \mathcal{B}_\psi \Gamma_{\Phi} \frac{\rho_\Phi}{M_\Phi} \, ,
\label{BE_psi}
\\
&&\dot{n}_S + 3 H n_S =  \mathcal{B}_S \Gamma_{\Phi} \frac{\rho_\Phi}{M_\Phi}  - \Gamma_{S} n_S  \, ,
\label{BE_b}
\\
&&
\dot{\rho}_{\hbox{\tiny SM}} + 4 H \rho_{\hbox{\tiny SM}} =  E_S\, n_S \, \Gamma_S , ,
\label{BE_rad}
\end{eqnarray}
where $\rho_\Phi=M_\Phi n_\Phi$ is the inflaton energy density, $n_\psi$ and $n_S$ are the dark fermion and dark scalar number density, $E_S$ is the energy of the dark scalar (cfr.~eq.~(\ref{energy_dark})) and $\rho_{\hbox{\tiny SM}}$ is the energy density of the SM radiation component. $\Gamma_\Phi$ is the total decay width of the inflation, namely the sum of the decay widths in eqs.~(\ref{decay_phi_to_psi}) and (\ref{decay_phi to_b}), the latter are expressed in terms of the corresponding branching ratios $\mathcal{B}_{\psi}$ and $\mathcal{B}_{S}$ in eqs.~(\ref{BE_psi}) and (\ref{BE_b}). The left-hand side of eq.~(\ref{BE_rad}) is obtained imposing the equation of state for the pressure and energy density of radiation  $p_{\hbox{\tiny SM}}=\rho_{\hbox{\tiny SM}}/3$. As in the usual Boltzmann approach, the terms on the right-hand side of each equation describe gain and loss terms for the given species due to the corresponding processes.\footnote{We checked that the $2 \to 2$ process $ss \to hh$ is negligible with respect to $s \to hh$ in the model parameter space. Therefore, we do not include it in the rate equations.}  We assume that the dark  scalars do not equilibrate with the SM plasma, which is a rather good approximation due to the very small portal couplings that get realized in our framework. 
The four components enter the Hubble rate as follows
\begin{equation}
    H= \left[ \left(\frac{1}{3 M_{\textrm{P}}^2}  \right)\left( \rho_\Phi + \rho_\psi + \rho_S + \rho_{\hbox{\tiny SM}} \right)  \right]^{\frac{1}{2}} \, ,
\end{equation} 
where we write the energy density stored in the dark particles as $\rho_\psi=E_\psi n_\psi$ and $\rho_S=E_S n_S$~\cite{Arcadi:2011ev}. The expression of the energy can be taken to be~\cite{Takahashi:2007tz,Dev:2013yza}
\begin{equation}
    E_{\psi(S)}(t) \simeq \sqrt{m^2_{\psi(S)}+\left[ \frac{M_\Phi \, \mathcal{B}_{\psi(S)}}{2} \frac{a(t_d)}{a(t)}\right]^2 } 
    \label{energy_dark}
\end{equation}
that holds for $m_\psi,m_S \ll M_\Phi$ and $t_d$ is the time when the inflaton decays, that we take to be the initial time when integrating the Boltzmann equations.\footnote{One could pursue a better treatment for $E_{\psi(S)}(t)$ at the injection from inflaton decays along with the discussion in ref.~\cite{Arcadi:2011ev}. However, we checked that the contribution of the three-momentum in (\ref{energy_dark}) is practically irrelevant for the numerics of our model and actually an approximation $\rho_\psi=m_\psi n_\psi$ and $\rho_S=m_S n_S$ works already pretty well. }

In order to solve the Boltzmann equations (\ref{BE_phi})-(\ref{BE_rad}), it is convenient to switch to another set of variables \cite{Chung:1998rq}
\begin{equation}
    \Phi = \frac{\rho_\Phi \, a^3}{M_\Phi} \, , \quad \Psi = n_\psi a^3 \, \quad \text{S} = n_S a^3 \, , \quad R = \rho_{\hbox{\tiny SM}} a^4 \, ,
    \label{def_var_Boltz}
\end{equation}
where one adopts the scale factor $a$ for the independent variable rather than the time. We define $A=a M_\Phi$, or equivalently $A=a/a_I$ with $a_I= 1/ M_\Phi$. 
Then, the Boltzmann equations (\ref{BE_phi})-(\ref{BE_rad}) become
\begin{eqnarray}
&&\frac{d \Phi}{d A}= - \frac{\sqrt{3}M_{\textrm{P}} \, }{M_\Phi^2} \frac{A^{\frac{1}{2}}  \Phi \, \Gamma_\Phi}{\left( \Phi + \frac{ E_S}{M_\Phi} \text{S}+ \frac{ E_\psi}{M_\Phi} \Psi + \frac{R}{A} \right)^{1/2}} \, ,
\label{BE_phi_bis}
\\
&&\frac{d \Psi}{d A}=  \frac{\sqrt{3}M_{\textrm{P}} }{M_\Phi^2} \frac{A^{\frac{1}{2}}  \Phi \, \mathcal{B}_\psi\, \Gamma_{\Phi}}{\left( \Phi + \frac{ E_S}{M_\Phi} \text{S} + \frac{ E_\psi}{M_\Phi} \Psi + \frac{R}{A} \right)^{1/2}} \, ,
\\
&&\frac{d \text{S}}{d A}=  \frac{\sqrt{3}M_{\textrm{P}}}{M_\Phi^2} \frac{A^{\frac{1}{2}} \left( \Phi \mathcal{B}_S \, \Gamma_{\Phi}  - \text{S} \, \Gamma_S \, \right)}{\left( \Phi + \frac{ E_S}{M_\Phi} \text{S} + \frac{ E_\psi}{M_\Phi} \Psi + \frac{R}{A} \right)^{1/2}} \, ,
\\
&&\frac{d R}{d A}=  \frac{\sqrt{3}M_{\textrm{P}} \, E_S }{M_\Phi^3}\frac{A^{\frac{3}{2}}  \text{S} \, \Gamma_S}{\left( \Phi + \frac{ E_S}{M_\Phi} \text{S} + \frac{ E_\psi}{M_\Phi} \Psi + \frac{R}{A} \right)^{1/2}} \, .
\label{BE_rad_bis}
\end{eqnarray}
We solve them with the initial conditions $\text{S}(A_I)=\Psi(A_I)=R(A_I)=0$, whereas the initial energy is entirely stored in the inflaton  $\Phi(A_I)=\rho_{\Phi,I}/M_\Phi^4$. As for the initial density of the inflaton, we use the estimation  $\rho_{\Phi,I} \approx M_\Phi^2 M_{\hbox{\scriptsize P}}^2$ that applies in the case of chaotic inflation. \begin{figure}[t!]
    \centering
    \includegraphics[scale=0.79]{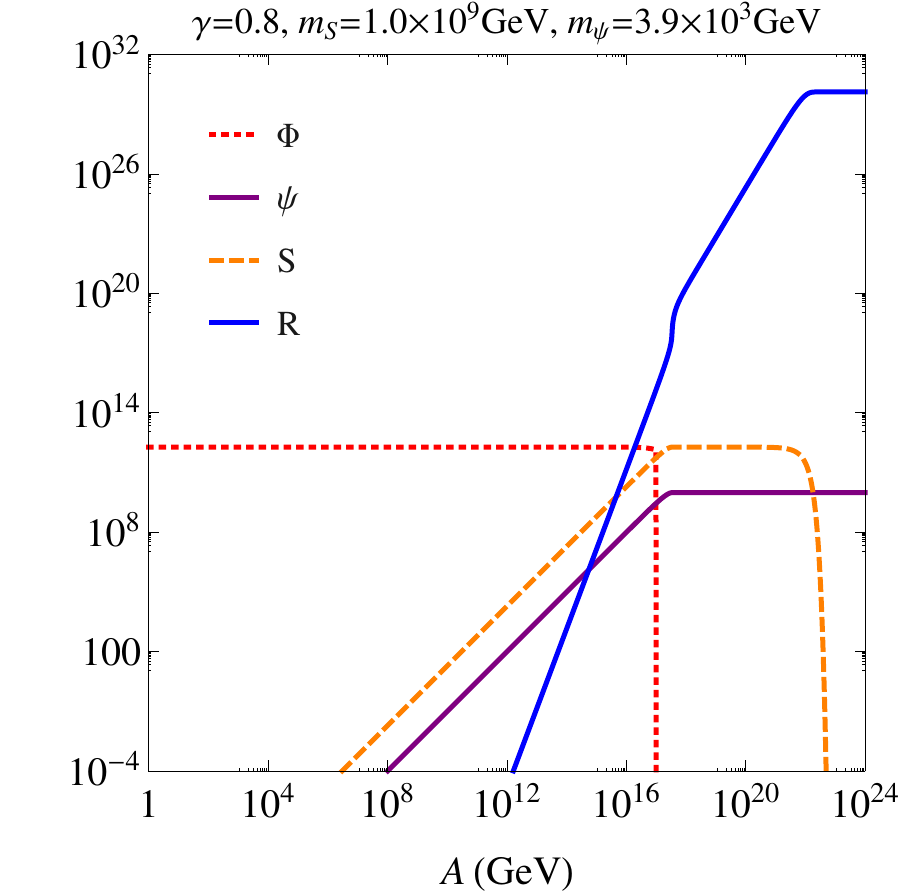}
    \hspace{0.3 cm}
    \includegraphics[scale=0.572]{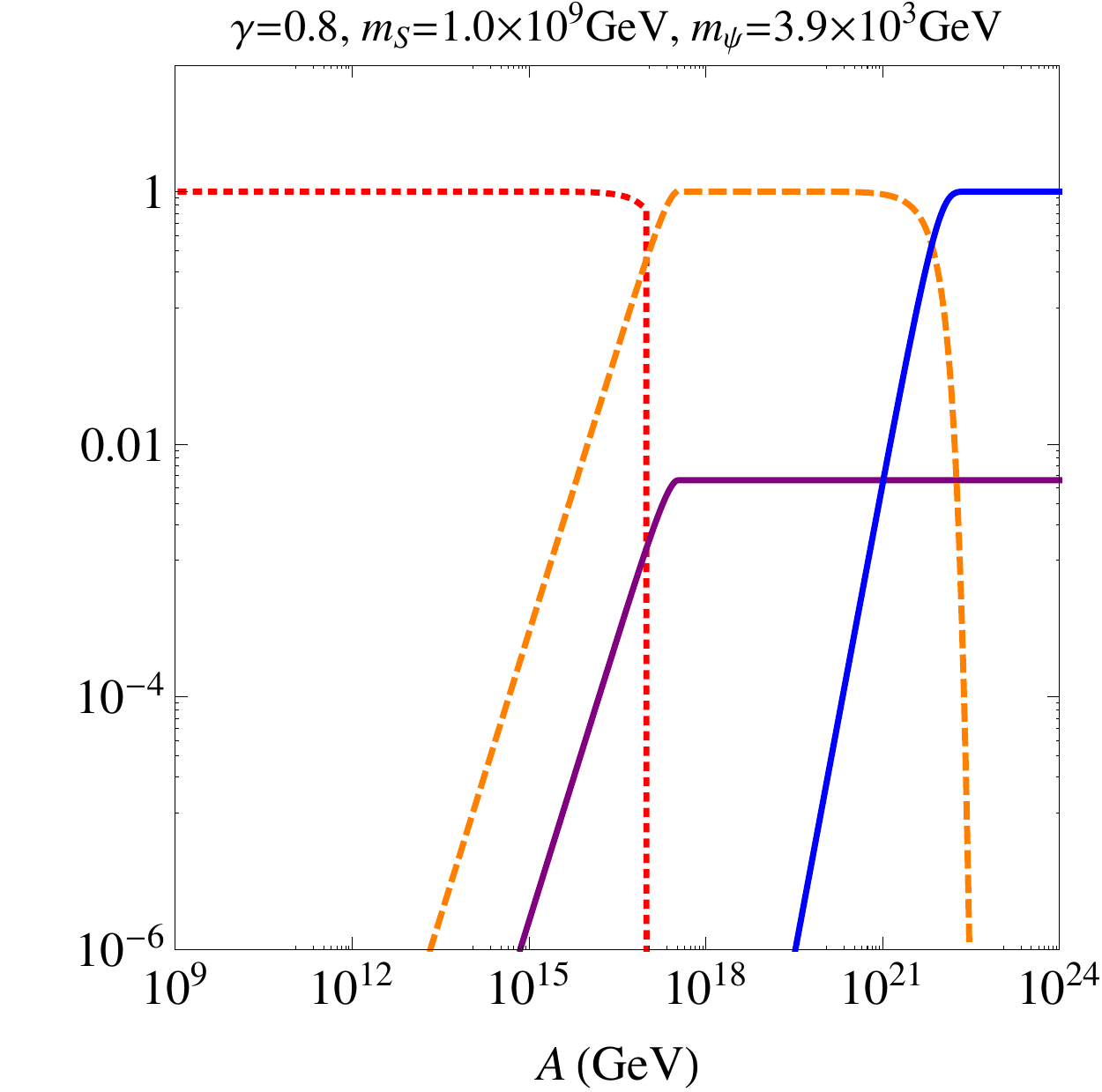}
    \caption{Solution of the Boltzmann equations (\ref{BE_phi_bis})-(\ref{BE_rad_bis}). The red-dotted line, purple-solid line, orange-dashed line and blue-solid lines correspond to the inflaton, dark fermion, dark scalar and SM radiation respectively. The dimension for $\Phi$, $\Psi$ and S is [GeV]$^3$, whereas for $R$ one has [GeV]$^4$. The right plot shows the same quantities, however rescaled and normalized as explained in the main text. We set $\gamma=0.8$ that gives $M_\Phi=8.9 \times 10^{12}$ GeV, then $m_S=1.0 \times 10^9$ GeV and $m_\psi=3.9 \times 10^3$ GeV.}
    \label{fig:gamma09_density}
\end{figure}

\begin{figure}[t!]
    \centering
    \includegraphics[scale=0.79]{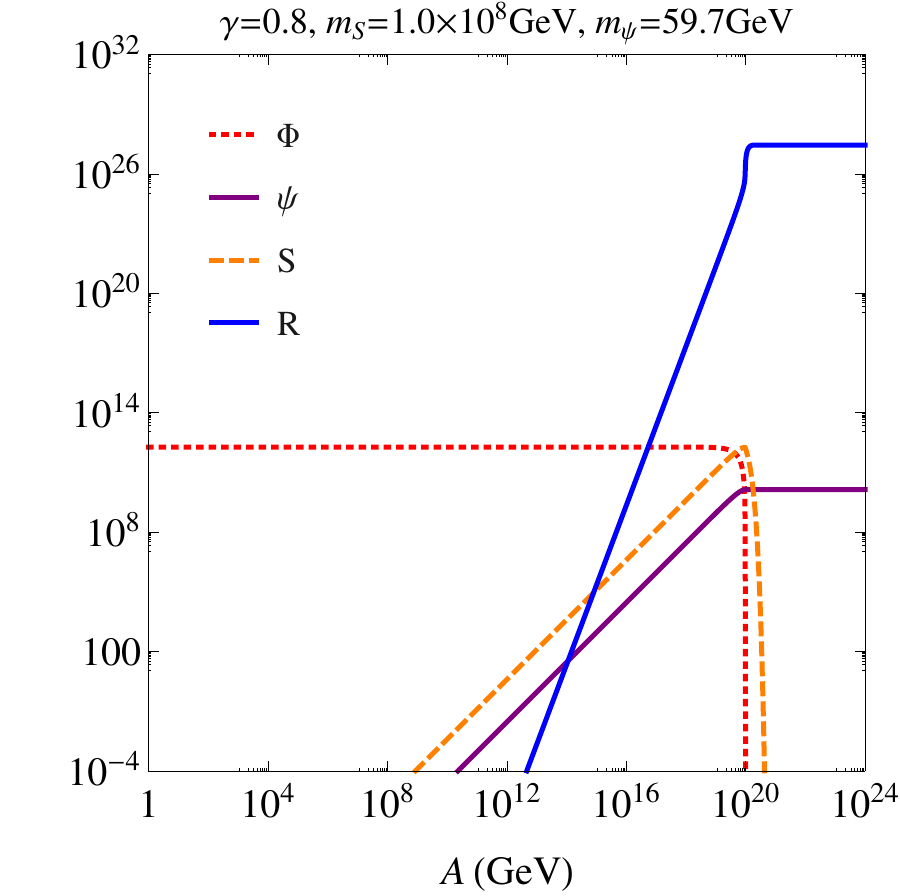}
    \hspace{0.3 cm}
    \includegraphics[scale=0.572]{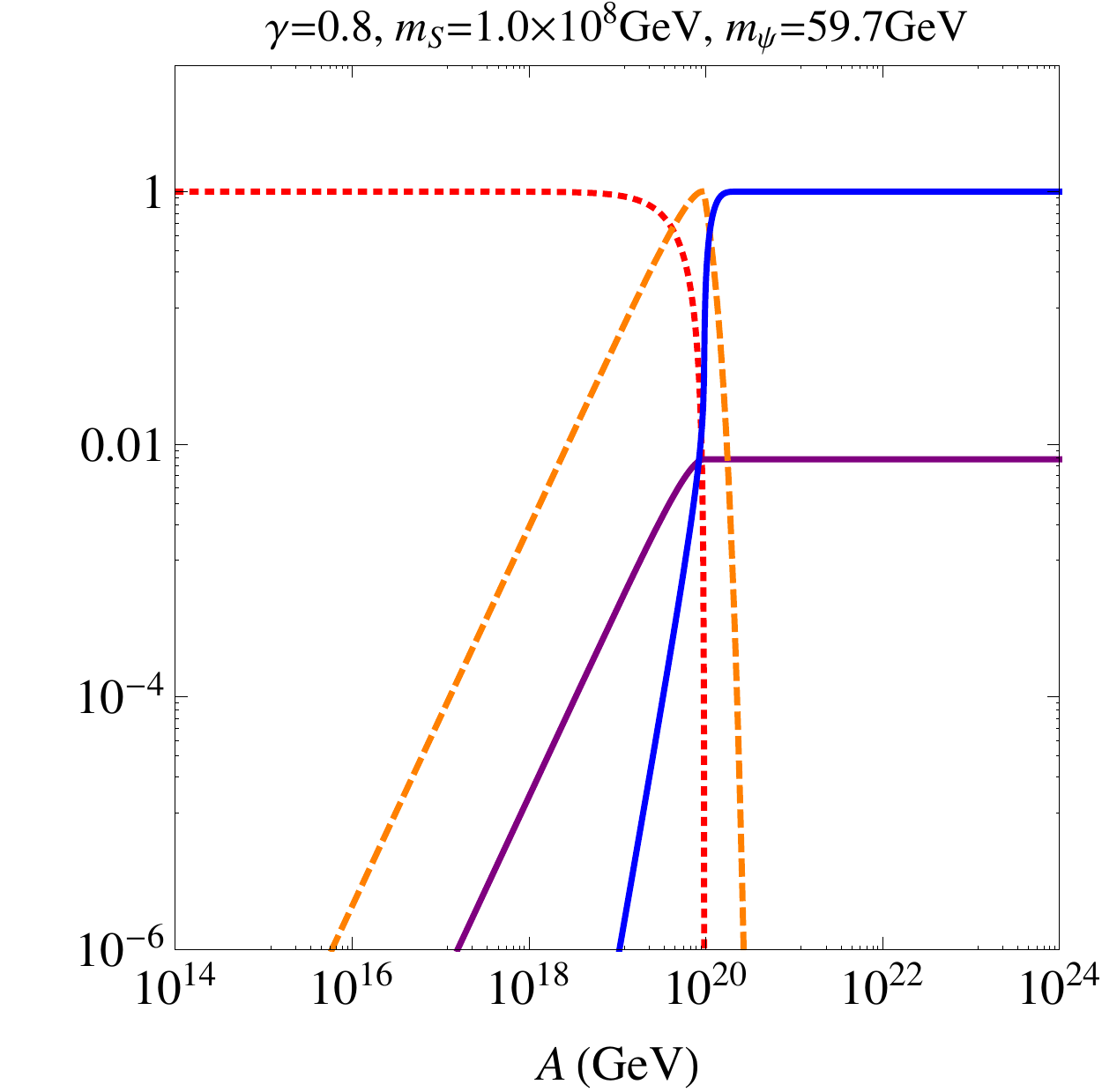}
    \caption{Color-style for the densities as in Fig.~\ref{fig:gamma09_density}. The dark particles have the following masses $m_S=10^8$ GeV and $m_\psi=59.7$ GeV.}
    \label{fig:gamma09_density_10_8}
\end{figure}

The results of the rate equations are collected in Figs.~\ref{fig:gamma09_density} and \ref{fig:gamma09_density_10_8} for two different choices of the model parameters $\beta_1$ and $\alpha$ that comply with the observed dark matter relic density, whereas  $\gamma=0.8$ is kept in both cases and it corresponds to $M_\Phi \simeq 8.9 \times 10^{12}$ GeV. In Fig.~\ref{fig:gamma09_density}, the mass of the dark scalar is $m_S=10^9$ GeV, whereas the mass of the dark matter fermion is $m_\psi=3.9 \times 10^3$ GeV. One may see that for a long time the inflaton density stays nearly constant (red-dotted line) and slowly populates the hidden sector with dark fermions and scalars (dark fermion and scalar particles in solid-purple and dashed-orange lines respectively). In the mean time, the dark scalar decays trigger the formation of the SM component, solid-blue line, which is smaller than the dark particles population for early times. The bulk of the inflaton decays happens for $A=10^{17}$ for this choice of the parameters, and one sees that the dark fermions and scalars freeze-in. Their later behaviour is different though. On the one hand, the dark fermion density stays constant ever since. On the other hand, the dark scalar exhibits a nearly constant density for a while, up until their population is very effectively depleted and no further decays can occur (this corresponds to the time when $\Gamma_{S} \siml H$). As a result, also the SM radiation density freezes to a constant value. In the right panel of Fig.~\ref{fig:gamma09_density}, we normalize the densities as follows: $\Phi/\Phi_I$, the $S$ scalar and $\Psi$ fermion densities are divided by the value of the dark scalar density at the onset of the freeze-in regime; finally the radiation is normalized by its constant value after $S$ decays no longer occur. For this choice of the parameters, the branching ratio of the process $\Phi \to \psi \psi$ is $\mathcal{B}_\psi \equiv \Gamma_{\Phi\to \psi \psi}/(\Gamma_{\Phi\to \psi \psi}+\Gamma_{\Phi\to ss}) \simeq 1.2 \times 10^{-3}$.

The relic density is fixed in terms of the relative density with the radiation component \cite{Chung:1998rq} \begin{equation}
    \frac{\Omega_{\hbox{\tiny DM}}h^2}{\Omega_{\hbox{\tiny SM}}h^2} = \frac{\rho_\psi (A_{\hbox{\tiny RH}})}{\rho_{\hbox{\tiny SM}} (A_{\hbox{\tiny RH}})} \frac{T_{\hbox{\tiny RH}}}{T_0} \, ,
\end{equation}
where $T_{\hbox{\tiny RH}}$ is the reheating temperature, $T_0=2.37 \times 10^{-13}$ GeV is the temperature today, and $\Omega_{\hbox{\tiny SM}}h^2=2.47 \times 10^{-5}$~\cite{Aghanim:2018eyx} is the present radiation energy density. This relation holds because the dark fermions freezes-in before the radiation density sets to a constant (see  solid purple and blue lines in Fig.~\ref{fig:gamma09_density} and \ref{fig:gamma09_density_10_8}).  Using the definitions (\ref{def_var_Boltz}) and in terms of the quantities extracted from the numerical solution of the Boltzmann equations (\ref{BE_phi_bis})-(\ref{BE_rad_bis}), we obtain  
\begin{equation}
    \Omega_{\hbox{\tiny DM}}h^2 = \frac{T_{\hbox{\tiny RH}}}{\hbox{GeV}} \frac{m_\psi}{M_\Phi} \frac{\Psi(A_{\hbox{\tiny RH}})}{R(A_{\hbox{\tiny RH}})} A_{\hbox{\tiny RH}} \, .
\end{equation}

The solutions of the Boltzmann equations for a second choice of the parameters is shown in Fig.~\ref{fig:gamma09_density_10_8}. The masses of the dark particles are $m_S = 10^8$ GeV and $m_\psi=59.7$ GeV that are compatible with the observed relic density. A smaller dark scalar mass has a relevant impact on the form of the solutions, as one may see by comparing the orange and blue lines in Fig.~\ref{fig:gamma09_density} and in Fig.~~\ref{fig:gamma09_density_10_8}. The main difference is that the dark scalar displays a constant density for a much smaller time and promptly decay as soon as the maximum abundance is reached. This is due to a larger decay width $\Gamma_S$ in eq.~(\ref{Higgs_width_1}) that follows from a smaller $m_S$ in the denominator and a larger portal coupling. Indeed, for a fixed value of $\gamma$, $\lambda_{HS}$ increases as the dark scalar mass decreases (see Fig.~\ref{fig:figMB_Higgs}). The branching ratio is  $\mathcal{B}_\psi  \simeq 4.2 \times 10^{-3}$ in this case.

A comment is in order for the non-thermally produced dark fermions. In our model, these particles are responsible for the present-day dark matter relic density. Therefore, they need to be non-relativistic (cold or warm) at the time of matter-radiation equality to allow structures formation. We use the comoving free-streaming length $\lambda_{\text{fs}}$ in order to assess the coldness of a non-thermal dark matter as explained in refs.~\cite{Takahashi:2007tz,Dev:2013yza}. From Lyman-$\alpha$ constraints, the comoving free streaming is bounded to be $\lambda_{\text{fs}} \siml 1$ Mpc. Since the dark fermions are produced in pairs from the inflaton decays, they can have large velocities. 
So one has to ensure that the dark fermions three-momenta are sufficiently red-shifted from the time scale of the inflaton decays to the matter-radiation equality. We find that the dark fermion belongs to the cold and warm categories as defined in ref.~\cite{Dev:2013yza} for the model parameters considered in Fig.~\ref{fig:gamma09_density} and Fig.~\ref{fig:gamma09_density_10_8} respectively. 
\begin{figure}[t!]
    \centering
    \includegraphics[scale=0.572]{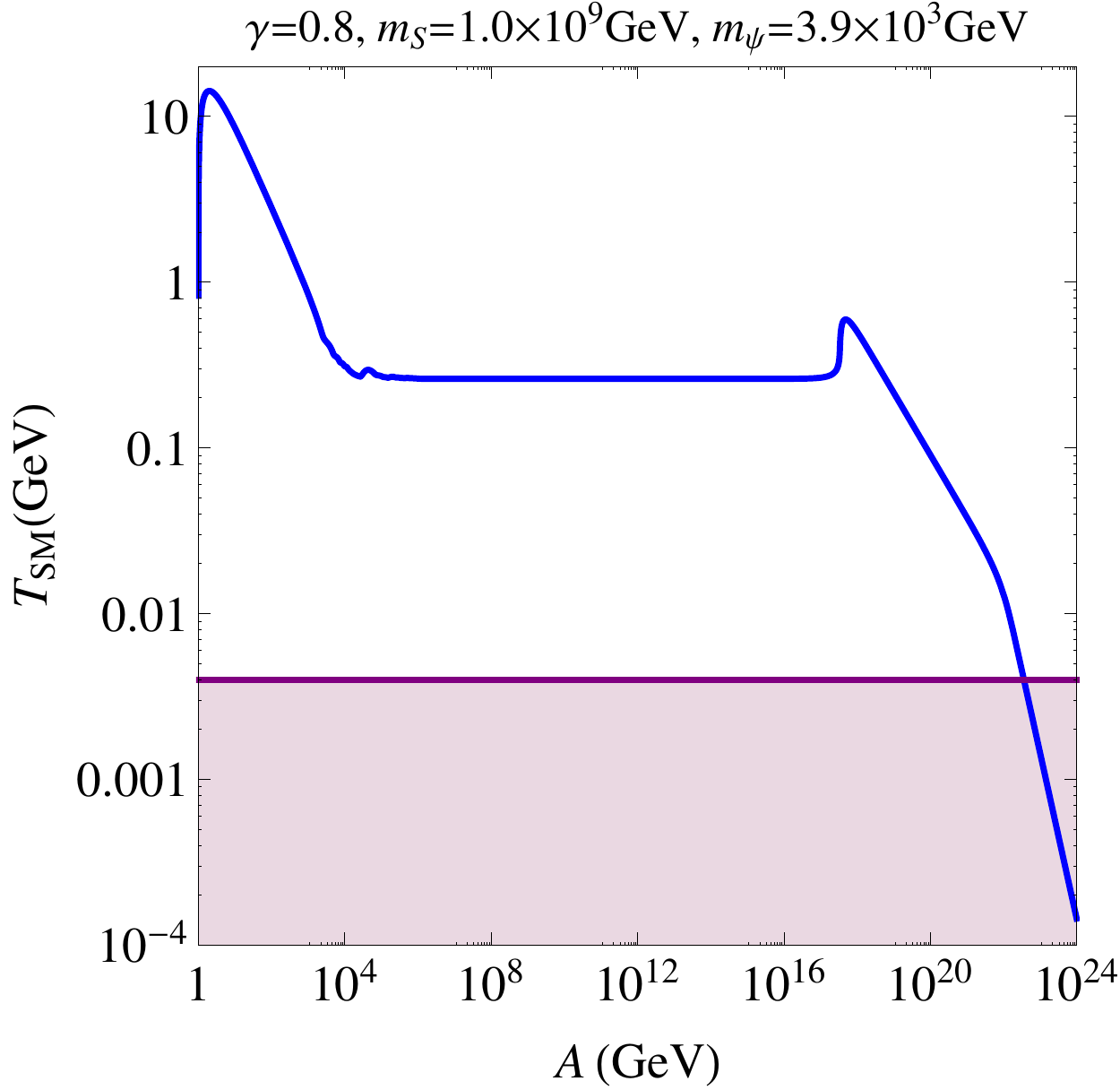}
    \hspace{0.25 cm}
    \includegraphics[scale=0.572]{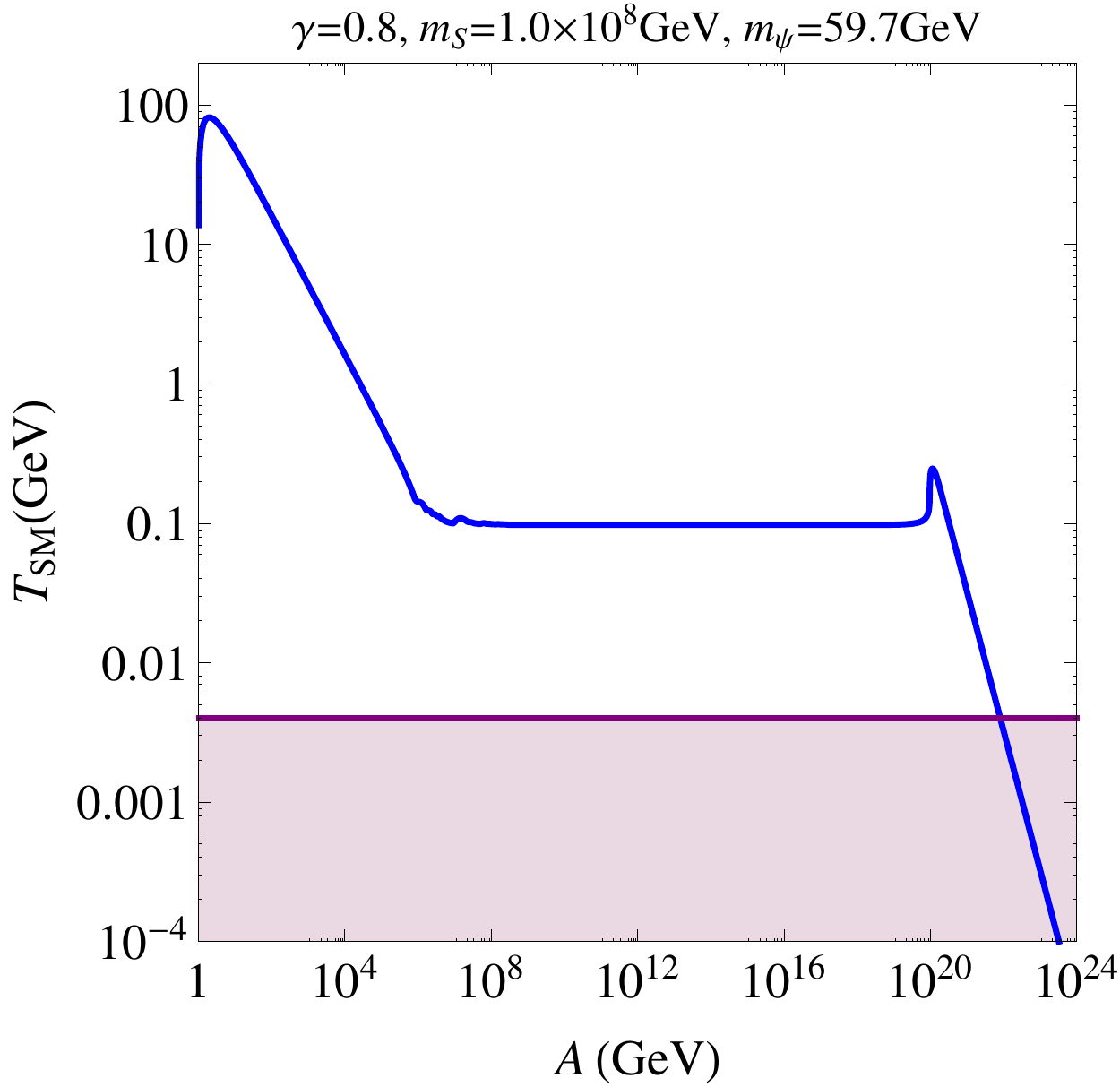}
    \caption{The SM temperature evolution as extracted from the solution of the Boltzmann equations is shown with solid blue line. The right and left panel correspond to the two different benchmark points discussed in the main text. The purple solid line indicates the lower bound of 4 MeV and the shaded area indicates smaller temperatures. Reheating has to happen above the $4$ MeV purple-solid line. }
    \label{fig:gamma09_temperature}
\end{figure}

Finally we discuss the temperature of the SM sector and the corresponding reheating. In Fig.~\ref{fig:gamma09_temperature}, the SM temperature is displayed for the two benchmark scenarios that also comply with a reheating temperature larger than 4 MeV. The temperature evolution can be obtained from the radiation density $R$ as follows
\begin{equation}
    T_{\hbox{\tiny SM}}= \left( \frac{30}{\pi^2 g_*( T_{\hbox{\tiny SM}})}\right)^{\frac{1}{4}} \frac{R^\frac{1}{4}}{A}M_\Phi \, ,
\end{equation}
where $R$ is the solution from the rate equations.
The plots show the evolution of the temperature along the different stages. First, as originally outlined in refs.~\cite{Chung:1998rq,Giudice:2000ex}, the reheating temperature is not the maximum temperature the thermal bath can reach. The very early rise up for the SM model temperature peaks roughly at 20 GeV (100 GeV) in the first (second) benchmark scenario, which is way larger than the reheating temperature at the onset of the $T_{\hbox{\tiny SM}} \propto a^{-1}$ scaling, that we find to be $T^{\hbox{\tiny RH}}_{\hbox{\tiny SM}} \simeq 19$ MeV ($T^{\hbox{\tiny RH}}_{\hbox{\tiny SM}} \simeq 182$ MeV). We extract the reheating temperature numerically and similarly to what it is done in ref.~\cite{Chung:1998rq,Giudice:2000ex,Arcadi:2011ev}. Basically, we match the $T_{\hbox{\tiny SM}} \propto a^{-3/8}$ behaviour obtained in the phase when the dark scalar acts as a large source of entropy to the   $T_{\hbox{\tiny SM}} \propto a^{-1}$ scaling in the subsequent radiation dominated regime, as clearly visible in the left plot in Fig.~\ref{fig:gamma09_temperature}. 
For the second choice of the parameters, the regime  $T_{\hbox{\tiny SM}} \propto a^{-3/8}$ is actually absent, and the radiation stage kicks in very sharply after the dark scalars reach their maximum density and they promptly decay into SM degrees of freedom. The second bumpy rise, that appears in both the cases, is due to the stage when the dark scalar decays are very effective and the abundance of dark scalars is entirely depleted, namely when $\Gamma_{S} \simeq H$.   

The numerical results can be compared with the instantaneous reheating temperature (\ref{SM_T_rh}) for the two benchmark scenarios, that read $T^{\hbox{\tiny RH},i}_{\hbox{\tiny SM}} \simeq 23$ MeV and 373 MeV respectively. Alternatively, one can obtain another estimate for the SM reheating temperature as outlined in ref.~\cite{Tenkanen:2016jic}, where the corresponding expression differs from that in (\ref{SM_T_rh}) by a factor $2^{5/4}$. In this latter case, one finds  10 MeV and 156 MeV.

\section{Conclusions and discussion}
\label{sec:conclusion}
In this paper we studied the post-inflationary dynamics of a CGUT model as given in (\ref{CFTSU(5)}). With respect to the original formulation, we include additional degrees of freedom that account for a hidden sector and the SM sector by respecting conformal invariance, see eq.~(\ref{bosf}). Dark fermions and dark scalars of the hidden sector are directly coupled to the inflaton, whereas the SM Higgs boson acts as a portal between the visible and the dark sector through a coupling shared with the dark scalars. 

As typical of conformal theories, no mass scale appears in the fundamental action (or Lagrangian). A non-vanishing VEV for the inflaton field is generated through radiative corrections induced by the GUT fields after the conformal and GUT symmetries are broken. The Planck mass is dynamically generated when the inflaton reaches its VEV. The same inflaton VEV is responsible for the masses of the dark particles. Moreover, the strength of the interactions between the inflaton and the dark particles is dictated by the form of the conformal couplings in (\ref{conf_coupl_matter}). To the best of our knowledge, this framework has not been implemented before within models that involve both conformal and GUT symmetry. 

A first orientation for the relevant parameter space of the model is given by the proton life time. In CGUT inflation, the proton life time can be extended beyond the observed limit $\tau_{p, \hbox{\scriptsize exp}} > 10^{34}$ yrs. This is a compelling guidance through the model parameter space and we restrict our analysis to the range where the model predicts $\tau_p \simg 60 \tau_{p, \hbox{\scriptsize exp}}$, where $\tau_{p, \hbox{\scriptsize exp}}$ is  the observed experimental bound (see Fig.~\ref{fig:protondecay}).   

In the present model, the electroweak symmetry breaking can be traced back to the inflaton VEV. Indeed, even though there is no direct coupling between $\phi$ and $H$ in the potential (\ref{conf_coupl_matter}), the dark scalar acts as a mass-scale generator from the inflaton sector to the SM domain. First, the VEV of the inflaton induces a mass term $\mu_S$ for the dark scalar, which triggers a spontaneous breaking of the dark scalar symmetric potential. As a result, a Higgs mass term $\mu_H$ is in turn induced, and the Higgs potential acquires its standard form as given in eq.~(\ref{higgs_0_pot}). Of course, in order to reproduce the observed Higgs mass, the model parameter has to be constrained accordingly. The main outcome is that very small portal couplings are consistent with the Higgs boson mass (cfr.~eq.~(\ref{higgsms}) and Fig.~\ref{fig:figMB_Higgs}). In our construction,  conformal symmetry breaking is at the origin for generating the relevant scales from inflation down to the electroweak scale.  

In section~\ref{sec:numerical_final}, we studied the evolution of the inflaton field, the dark particles and the SM radiation that are respectively originated directly and indirectly from the inflaton decays. The dark fermions are inert stable particles, they account for the present day relic density and they are produced non-thermally by the inflaton decays. The dark scalars, also produced non-thermally from the inflaton, are unstable and decay in turn into SM Higgs bosons. This way, the energy stored in the inflaton field can also leak to the SM, even though there is no direct interactions among the inflaton and the SM. The model at hand can accommodate the observed relic density and a reheating temperature compatible with the BBN requirements. In solving the Boltzmann equations, we track the SM temperature along with the whole post-inflationary dynamics. We extract the reheating temperature numerically and, for the two benchmarks choices considered in this work, we are able to reproduce the observed relic density of dark matter and a reheating temperature $T^{\hbox{\tiny RH}}_{\hbox{\tiny SM}} \simg 4$ MeV. We focused on $\gamma=0.8$ ($M_{\Phi} \simeq 8.9 \times 10^{12}$ GeV) that provides a broad window to allow for sufficiently high reheating temperatures. A more detailed study of the whole parameter space is beyond the scope of the present paper. In our model, it is likely that the hidden sector does thermalize and does not have its own temperature. We assumed that this holds in any stage of the evolution. 

It is not obvious to realize both the observed relic density and a reheating temperature larger than $4$ MeV in our framework. From the analysis given in section~\ref{sec:numerical_final}, we see that larger reheating temperatures (smaller dark scalar masses) push the dark fermion mass to very small values. Given the lower bound on the dark fermions mass $m_\psi >1$ GeV as obtained from the inflaton decay (see Fig.~\ref{fig:figMass} and section~\ref{sec:dark_masses}), a viable dark matter candidate cannot be provided for too high reheating temperatures. On the other hand, increasing the dark scalar masses implies having a less efficient energy transfer to the SM sector (because $\Gamma_{S}$ becomes smaller), that results in too low reheating temperatures.  

As far as the model is implemented in this work, the dark particles are rather hard to be searched for at experimental facilities. The dark scalars are heavier than $10^3$ TeV and couple feebly with the SM Higgs boson $\lambda_{HS}\siml 10^{-7}$. The dark fermions do not interact with any other field but the inflaton. Therefore, direct, indirect and collider searches cannot be applied for our dark matter candidate. It is possible to consider couplings between the dark fermion and the SM Higgs boson whenever we allow for a terms $\bar{\psi} \psi S$ that respect the conformal invariance of the potential. For small dark matter masses, this might open up a production reaction through $s$-mediated processes like $hh \to \bar{\psi} \psi$~\cite{Chung:1998rq,Giudice:2000ex,Gelmini:2006pw,Arcadi:2011ev}, that can also give viable channels for detection strategies of the dark matter candidate. We leave the inclusion and assessment of detectable dark matter signatures, together with gravitational waves production from the breaking of conformal, GUT and dark sector symmetries for future research on the subject.

\section*{Acknowledgements}
We thank Giorgio Arcadi for useful discussions, and Anish Ghoshal for reading the manuscript and useful comments. K.S.K is supported by the Netherlands Organization for Scientific Research (NWO) grant number 680-91-119. 
\appendix
\numberwithin{equation}{section}

\section{Dark scalar and Higgs boson mixing}
\label{sec:scalar_mix}
We give here some details for scalar mixing at zero temperature. We adapt the derivation given in ref.~\cite{Kahlhoefer:2015bea}. Considering the SM Higgs  and the dark scalar, the most general scalar potential after electroweak and $Z_2$ dark symmetry breaking can be written as
\begin{equation}
    V_{s,h}=-\frac{\mu_S^2}{2}(v_S+s)^2 + \lambda_S  (v_S+s)^4 -\frac{\lambda_{HS}}{4}  (v_S+s)^2(v_H+h)^2 +\lambda_H (v_H+h)^4 \, ,
\end{equation}
where we notice the absence of a mass term $\mu_H$ from the start for the Higgs boson. This is instead given by the portal coupling and the VEV of the dark scalar. In the limit $\lambda_{HS}=0$ we obtain
\begin{equation}
    v_S^2=\frac{\mu_S^2}{\lambda_S} \, , \quad  m_S^2 = 2 \lambda_{S} v_S^2  \, ,
\end{equation}
and a vanishing Higgs VEV,  $v_H =0$. With a non-vanishing portal coupling, one then obtains for the minimum ($4 \lambda_H \lambda_S > \lambda_{HS}$)
\begin{eqnarray}
v_H^2&=&\frac{2 \lambda_{HS} \, \mu_S^2}{4 \lambda_H \lambda_S -\lambda_{HS}^2} \, , 
\\
v_S^2&=&\frac{4\lambda_{H} \, \mu_S^2}{4 \lambda_H \lambda_S -\lambda_{HS}^2} \, .
\end{eqnarray}
The mass squared eigenvalues read
\begin{equation}
    m^2_{1,2}= \lambda_H v_H^2 + \lambda_S v_S^2 \, \mp \, \sqrt{(\lambda_S v_S^2-\lambda_H v_H^2)^2+\lambda_{HS}^2 v_S^2 v_H^2} \, ,
\end{equation}
that correspond to the scalar fields eigenstates 
\begin{eqnarray}
    H_1= s \sin{\theta} +h \cos{\theta} \, , 
    \\
    H_2=s \cos{\theta}-h \sin{\theta} \, .
\end{eqnarray}
and tangent of twice the mixing angle is given by 
\begin{equation}
    \tan 2 \theta = \frac{\lambda_{HS} \, v_S v_H}{\lambda_S v_S^2 - \lambda_H v_H^2} \, .
\end{equation}
In the small mixing limit $\lambda_{HS} \ll \lambda_H, \lambda_S$, one expands the mass eigenvalues expression and obtains 
\begin{equation}
    m_1 \approx 2 \lambda_H v_H^2 \equiv m_H^2 \, \quad m_2^2= 2 \lambda_S v_S^2 \equiv m_S^2 \, ,
\end{equation}
and accordingly the mixing angle becomes (neglecting higher order in the portal coupling)
\begin{equation}
    \theta \approx \frac{\lambda_{HS} \, v_S v_H}{m_S^2-m_H^2} \, .
\end{equation}

\section{Doublet-triplet splitting and CGUT}
\label{sec:doublet_triplet}
In the action (\ref{CFTSU(5)}) we neglected the couplings between $\phi$ and $H_5$, that comprises the Higgs doublet and the triplet-color Higgs, as well as the couplings between the adjoint GUT Higgs field $\Sigma$ and $H_5$. In so doing, we obtained a simplified version of the action that still allows us to generate the inflaton VEV from radiative corrections and reheating of the SM through the dark scalar. In this appendix, we discuss these couplings in relation to the doublet-triplet splitting problem in SU(5). Since our principle is conformal symmetry, we write here the following terms in the potential 
\begin{equation}
V( \phi,\,\Sigma,\,H_5 ) = \gamma_1 f\LF\frac{\phi}{\chi}\RF \phi^2 H_5^\dagger H_5 -  \gamma_2 H_5^\dagger \Sigma^2 H_5+ \lambda_{H_5}\LF H_5^\dagger H_5 \RF^2.
\label{V_split_1} 
\end{equation}   
The field $H_5\equiv \LF H_3,\,H_2 \RF ^\text{T}$ contains the colour triplet $H_3$, the  SM electroweak doublet $H_2$, and no mass scale appears in the potential (\ref{V_split_1}). Assuming the GUT symmetry is broken by (\ref{GUTfieldVEV-1}) and substituting (\ref{sigma2field}), we express the potential in terms of $H_3$ and $H_2$ and as
\begin{equation}
\begin{aligned}
V( \phi,\,\Sigma,\,H_5 ) = & f\LF \frac{\phi}{\chi} \RF \phi^2 H_3^\dagger \LF \gamma_1-\frac{2\lambda_2\gamma_2}{15\lambda_c} \RF H_3+ f\LF \frac{\phi}{\chi} \RF \phi^2 H_2^\dagger \LF \gamma_1-\frac{3\lambda_2\gamma_2}{10\lambda_c} \RF H_2 \\ &+ \lambda_{H_5} \LF H_3^\dagger H_3+H_2^\dagger H_2 \RF^2.
\end{aligned}
\end{equation}
When the field $\phi$ and $\sigma$ reach their VEVs, $H_3$ and $H_2$ acquire a mass terms each. In GUT frameworks, one aims at a light Higgs doublet (at the weak scale) and much heavier coloured Higgs triplet (to suppress the proton decay). To achieve this we require  
\begin{equation}
\gamma_1 \approx \frac{3\lambda_2\gamma_2}{10\lambda_c} \, .
\end{equation}
With this condition, the doublet become nearly massless and triplet acquire a (physical) mass
\begin{equation}
 M_{H_3}^2= \frac{\lambda_2\gamma_2}{6\lambda_c} \gamma^2 v_\phi^2 =\frac{\lambda_2\gamma_2}{ \lambda_c} M^2_{\textrm{P}}(1-\gamma^2) \, .
\end{equation}
We understand the coloured Higgs triplet potential with a positive mass term and, therefore, the corresponding potential does not develop a VEV.  
Our implementation is similar to the one presented in ref.~\cite{Ellis:2014dxa}.

\bibliographystyle{JHEP.bst}
\bibliography{InflDM.bib}

\end{document}